\documentclass{aastex}
\usepackage{apj}
\shortauthors{Macri {\it et al.}}
\shorttitle{Cepheid distance to \n2841}
\begin{document}
\def \muv  {31.23\pm0.08}
\def \mui  {30.96\pm0.06}
\def \mvi   {0.27\pm0.03}
\def \muoi {30.58\pm0.06}
\def \muoe {30.74\pm0.23}
\def \dist {14.1\pm1.5}
\def \ncv{79}
\def \nr{50}
\def \nv{29}
\def \nc{26}
\def \nm{3}
\def \ng{18}
\def \vi{V-I}
\def \bv{B-V}
\def \evi{E(\vi)}
\def \ebv{E(\bv)}
\def \n2841{NGC$\,$2841}
\def \sn {SN$\,$}
\def \Deg{${}^\circ$\llap{.}}
\def \Sec{{}^{\prime\prime}\llap{.}}
\def \m{$-$}

\title{The Discovery of Cepheids and a New Distance to \n2841 \\
Using the Hubble Space Telescope$^\dagger$}

\author{L.M.~Macri\altaffilmark{1}, P.B.~Stetson\altaffilmark{2},
G.D.~Bothun\altaffilmark{3}, W.L.~Freedman\altaffilmark{4}, \\
P.M.~Garnavich\altaffilmark{5}, S.~Jha\altaffilmark{1},
B.F.~Madore\altaffilmark{4,6} \& M.W.~Richmond\altaffilmark{7}}

\altaffiltext{1}{Harvard-Smithsonian Center for Astrophysics, 60 Garden St.,
Cambridge, MA 02138, USA. lmacri@cfa.harvard.edu}

\altaffiltext{2}{Dominion Astrophysical Observatory, Herzberg Institute of
Astrophysics, National Research Council of Canada, 5071 West Saanich Road,
Victoria, BC V8X 4M6, Canada.}

\altaffiltext{3}{Department of Physics, University of Oregon, 1371 E 13th Ave.,
Eugene, OR 97403, USA.}

\altaffiltext{4}{Observatories of the Carnegie Institution of Washington, 813
Santa Barbara St., Pasadena, CA 91101, USA.}

\altaffiltext{5}{Physics Department, 225 Nieuwland Science Hall, University of
Notre Dame, Notre Dame, IN 46556, USA.}

\altaffiltext{6}{Infrared Analysis and Processing Center, California Institute
of Technology, 770 South Wilson Avenue, Pasadena CA 91125, USA.}

\altaffiltext{7}{Physics Department, Rochester Institute of Technology,
1 Lomb Memorial Dr., Rochester, NY 14623}

\begin{abstract}
We report on the discovery of Cepheids in the spiral galaxy \n2841, based on
observations made with the Wide Field and Planetary Camera 2 on board the {\it
Hubble Space Telescope}. \n2841 was observed over 12 epochs using the F555W
filter, and over 5 epochs using the F814W filter. Photometry was performed
using the DAOPHOT/ALLFRAME package.

We discovered a total of $\nv$ variables, including $\ng$ high-quality
Cepheids with periods ranging from 15 to 40 days. Period-luminosity relations
in the V and I bands, based on the high-quality Cepheids, yield a
reddening-corrected distance modulus of $\muoe$~mag, which corresponds to a
distance of $\dist$ ~Mpc. Our distance is based on an assumed LMC distance
modulus of $\mu_0=18.50 \pm 0.10$~mag ($D=50\pm 2.5$~kpc) and a metallicity
dependence of the Cepheid P-L relation of $\gamma_{VI}=-0.2\pm0.2$~mag/dex.
\end{abstract}


\renewcommand{\thefootnote}{\fnsymbol{footnote}} 

\footnotetext[2]{Based on observations with the NASA/ESA {\it Hubble
Space Telescope}, obtained at the Space Telescope Science Institute,
operated by AURA, Inc. under NASA contract No. NAS5-26555.}

\renewcommand{\thefootnote}{\arabic{footnote}}

\vspace{-12pt}

\section{Introduction}

This paper presents the discovery of $\nc$ Cepheids and a distance
determination to \n2841. \n2841 is an isolated, flocculent spiral
galaxy, classified as Sb(r): I \citep{RC3} or Sb I \citep{RSA}, and
located at $\alpha = $ 9$^{\rm h}$ 22$^{\rm m}$ 03$^ {\rm s}$, $\delta
=$ +50$^{\circ}$ 58${\arcmin}$ 36${\arcsec}$ (J2000.0). It exhibits a
heliocentric redshift of $+683$~km/s, and has a major axis diameter of
$8\farcm 3$. \citet{NBG} places it in the northwestern-most
association of the Leo Spur (Group 15+10).

The inclination angle of \n2841 is $64^{\circ}$ \citep{RC3}, making it a
suitable calibrator of the Tully-Fisher relation. Furthermore, the large
observed 21-cm line-width of \n2841 \citep[$614\pm4$~km/s,][]{ro00}
significantly extends the range in rotation velocity over which the
Tully-Fisher relation has been calibrated using Cepheid distances. This topic
will be covered by \citet{bo01}, which will also address the implications of
the \n2841 Cepheid distance regarding the predictions from the MOND (MOdified
Newtonian Dynamics) theory.

\n2841 has been the host of four supernovae in the last century: \sn1912A,
\sn1957A, \sn1972R and \sn1999by. The last one is a well-observed type Ia SN
which has been classified as a fast decliner (hence sub-luminous). Thus, it is
of special importance for the calibration of the type Ia SNe distance
indicator.  This topic will be addressed by \citet{ga01}.

\S 2 describes the observations and preliminary reductions of the images. \S 3
contains the details of photometry and calibration of the instrumental
magnitudes. \S 4 presents the sample of variables and their properties. \S 5
describes the fiducial Cepheid Period-Luminosity relations used in our
analysis, derives a distance modulus, and lists the sources of uncertainty. \S
6 discusses some implications of our distance measurement, and \S 7 presents
our conclusions.

\section{Observations and Reductions}

\n2841 was observed by the Wide Field and Planetary Camera 2 (WFPC2)
\citep{bir00} on board HST on twelve epochs between 2000 February 29 and April
19.  The observations were performed using the F555W (approximately Johnson
V) and the F814W (approximately Kron-Cousins I) filters. All twelve epochs
contained a pair of CR-split images in the F555W band, while five epochs
contained an additional pair of CR-split images in the F814W. Each individual
image had an exposure time of 1100s. All observations were made with the
telescope guiding in fine lock with a stability of approximately 3 mas. The
gain and readout noise were 7 e$^{-}$/DN and 7 e$^{-}$, respectively. The CCD
was operated at a temperature of --88$^\circ\,$C for all observations. A log of
the observations is presented in Table~1. The sampling strategy followed in
these observations was similar to the one employed by the {\it HST} Key Project
on the Extragalactic Distance Scale \citep{fre01}, It followed a power-law
distribution in time, which provides an optimum sampling of the light curves of
Cepheids in the period range between 10 and 50 days. 


\clearpage

The images were calibrated using the pipeline processing at the Space Telescope
Institute (STScI). The full reduction procedure \citep{hol95a} consisted of: a
correction for A/D conversion errors, the subtraction of a bias level for each
chip, the subtraction of a superbias frame, the subtraction of a dark frame, a
correction for shutter shading, and a division by a flat field. Furthermore,
the images were corrected for vignetting and geometrical distortions in the
WFPC2 optics (the latter correction was done using files kindly provided by
J.~Holtzman). Lastly, the images were multiplied by four and converted to
two-byte integers, to reduce disk usage and allow image compression. This
conversion led to an effective readout noise of 4 DN and a gain of
1.75e$^-$/DN.

Figure~1 displays a ground-based image of \n2841, obtained at the Whipple
Observatory 1.2-m telescope, with a super-imposed mosaic of the WFPC2 field of
view. The mosaic is shown in greater detail in Figure~2.

\section{Photometry and Calibration}

Photometry of \n2841 was performed using the DAOPHOT/ALLFRAME package.
\citet{ste94,ste98} contain comprehensive overviews of the package and the
reduction strategy, which we briefly summarize here.

As in the case of previous ALLFRAME reductions of WFPC2 data, we used external
point-spread functions (PSFs), determined by P.B.S. for each chip and filter
combination from high $S/N$, uncrowded observations of Galactic globular
cluster fields. These external PSFs take into account the variation in the PSF
across the field of view of each WFPC2 chip.

The photometric reduction started with a preliminary detection of sources in
each individual image. DAOMATCH and DAOMASTER were used to derive offsets and
rotations between frames, relative to the first F555W frame. In the case of
these observations, the shifts were $<1$~pixel for 31 frames, and $\sim2$
~pixels for the other two. Once the transformations were established, MONTAGE
was used to create a medianed image, free of cosmic rays and chip defects. The
FIND routine in DAOPHOT was used to detect the sources present in that image,
and their positions were refined by running ALLSTAR. The resulting list of
objects was used by ALLFRAME to perform the photometry of each object in every
frame.

The conversion from PSF to standard \citep{hol95b} 0.5$\arcsec$-radius
aperture magnitudes involved the determination of growth curves.  First, we
selected bright and isolated stars (hereafter, ``secondary standards'') that
were located in areas with low and slowly-varying backgrounds. Next, all other
objects were removed from the frames and aperture photometry was carried out
at a variety of radii. DAOGROW \citep{ste90} was used to derive growth curves
and to obtain $0.5\arcsec$-radius aperture magnitudes for the secondary
standards in each frame. Lastly, we used COLLECT to calculate the ``aperture
correction'' coefficient for each frame from the difference between PSF and
aperture magnitudes of the secondary standards.

The conversion from $F555W$ and $F814W$ magnitudes to Johnson
V and Kron-Cousins I bandpasses followed the precepts of \citet{hol95b}:

\ \par

\begin{eqnarray}
V&=&F555W-0.052 [\pm0.007]\ (\vi)\\ \nonumber
 & & + 0.027 [\pm0.002]\ (\vi)^2+Z_{i,F555W}\\[6pt]
I&=&F814W-0.062 [\pm0.009]\ (\vi)\\ \nonumber
 & & + 0.025 [\pm0.002]\ (\vi)^2+Z_{i,F814W}
\end{eqnarray}

\vskip 3pt

\noindent{where $F555W$ and $F814W$ equal $-2.5\log({\rm DN\ s}^{-1})$ and the
color terms come from Table~7 of \citet{hol95b}. The zeropoint terms ($Z_i$),
which are listed in Table~2, were determined by P.B.S. and reflect evolution in
the understanding of the WFPC2 calibration since the publication of
\citet{ste98} (cf. Table~5 of that paper). For those epochs without
F814W data, the transformation of the $F555W$ magnitudes into the standard V
bandpass was carried out using the mean $\vi$ color of each object.}

Additionally, the magnitudes of each object were corrected for the effects of
charge transfer inefficiency following the prescription of \citet{ste98}. In
order to facilitate the comparison of our magnitude system with future studies,
we list in Table~3 the mean V and I magnitudes for the secondary standards
present in each chip.

\section{The Cepheids found in \n2841}

The initial search for variables was carried out using TRIAL, a program that
calculated the value of the $J$ statistic for each star, based on the V-band
data. The $J$ statistic is a robust index of variability developed by
\citet{ste96}, and it is designed so that objects with constant magnitudes
yield values near zero. The mean value and {\it r.m.s.} deviation of this index
for all stars present in our database was $0.01\pm0.29$. We flagged $\ncv$
objects with $J>0.8$ as possible variables, and extracted their individual
light curves from the photometry database for further analysis using a suite of
programs developed by the DIRECT team \citep{ka98}.

We calculated periods for the candidate variables using a modified version of
the Lafler-Kinman technique \citep{ste96}; we restricted the range of possible
periods to the range between 10 and 50 days. Next, template Cepheid light
curves \citep[derived by][]{ste96} were simultaneously fitted to the phased
V- and I-band photometry of each candidate. The use of templates allows a
robust classification of a candidate variable as a Cepheid, refines the period
of the variable, and yields reliable mean magnitudes through numerical
integration of the best-fit template. If the candidate object was classified as
a Cepheid, an iterative $3\sigma$ rejection algorithm was applied to the data
to reject anomalous photometric measurements, most of them associated with
cosmic-ray hits, and the light curve was analyzed again. About 5\% of the
individual data points were rejected in this manner.

The automated classification software rejected $\nr$ candidates and classified
$\nc$ objects as Cepheids. An additional $\nm$ objects are clearly variables,
and most likely Cepheids, but their periods are longer that our observing
window and we cannot obtain accurate values for their mean magnitudes.  Table~4
lists the following properties of the $\nv$ variables which passed the
analysis: identification number (C01-C$\nc$ in order of descending period, for
the Cepheids, and V01-V0$\nm$ in order of increasing R.A., for the other
variables); chip where the object is located; $(x,y)$ coordinates (based on the
coordinate system of the first V band image); J2000.0 coordinates; best-fit
period and uncertainty (Cepheids only); V and I mean magnitudes, derived from
the best-fit template light curves; and a selection flag (see \S5.2 for
details).

Figure~3 contains chip-wide images of each of the WFPC2 CCDs, indicating the
locations of the variables. Finding charts for each of the stars are displayed
in Figure~4; the charts encompass a 50 by 50 pixel area around each variable
(i.e., $2\Sec15\times 2\Sec15$ for the PC and $5\arcsec\times 5\arcsec$ for the
WF). The individual V and I band photometric measurements are listed in
Tables~5 and 6, respectively, and shown as phase-magnitude plots (for the
Cepheids) and as time series (for the other variables) in Figure~5. That Figure
also shows the V and I best-fit template light curves for each Cepheid, which
were used to determine their mean magnitudes.

Figure~6 shows a color-magnitude diagram of the $\sim 9000$ stars detected in
our images. As expected, the Cepheids occupy a region of the diagram that is
consistent with the location of the instability strip. Figure~7 shows the
observed differential luminosity functions for the V and I data, which indicate
that our completeness limits are V$\sim 26.5$~mag and I$\sim25.5$~mag.
Additionally, the I band luminosity function does not exhibit the signature
associated with the Tip of the Red Giant Branch \citep[c.f.~Figure 2 of][]
{szk00}, yielding a lower limit for the distance to \n2841 of $\sim29.5$~mag
or $\sim8$~Mpc.

\section{Period-Luminosity Relations \\ and Distance Moduli}
\subsection{Methodology}
The method used to derive V- and I-band apparent distance moduli is the
same one that was used by the {\it HST} Key Project on the Extragalactic
Distance Scale, which is described in detail by \citet{fre01}. It relies on
fiducial Period-Luminosity relations of LMC Cepheids, corrected for reddening
and scaled based on an assumed true distance modulus of $\mu_{0,{\rm LMC}}$ =
18.50$\pm$0.10~mag. The P-L relations used are those from \citep{ud99}, based
on a sample of $\sim 650$ OGLE Cepheids with periods ranging from 2.5 to 28
days:

\vskip -6pt
\begin{eqnarray}
M_V&=&-2.760(\pm0.03)\left[\log{P}
      -1\right] -4.218(\pm0.02), \\
M_I&=&-2.962(\pm0.02)\left[\log{P}
      -1\right] -4.904(\pm0.04).
\end{eqnarray}
\vskip 3pt

These relations can be combined to obtain an equation for the
reddening-corrected distance modulus of a Cepheid \citep{fre01}:

\vskip -6pt
\begin{equation}
\mu_0 =  W + 3.255(\pm0.01)\left[\log{\rm P} -1\right] +5.899(\pm0.01),
\end{equation}
\vskip 3pt

\noindent{where $W=V-2.45 (V-I)$ is the reddening-free Wesenheit magnitude
\citep{ma82}.}

\subsection{Distance moduli}

Before determining distance moduli from Equations 3-5, we must impose a cut in
period at the short end to avoid an incompleteness bias \citep[see Appendix A
of][]{fre01}. The completeness limits determined at the end of \S 4 imply that
this cut must be applied around $P=20$~d, resulting in the rejection of
Cepheids C01-C05. We also rejected Cepheids C07-C09 due to their rather poor
quality I band light curves. Thus, the final sample of Cepheids used for
distance determination consists of $\ng$ variables, identified with an asterisk
in the last column of Table~4.

Figures~8 and 9 show the V and I band P-L relations of the final sample of
Cepheids. By fitting the P-L relations described in Equations (3) and (4), we
obtain apparent V and I band distance moduli of $\mu_V=\muv$~mag and
$\mu_I= \mui$~mag (internal errors only), and a mean color excess of $\evi=
\mvi$~mag. The application of Equation (5) results in an extinction-corrected
distance modulus of $\mu_0=\muoi$~mag (internal errors only).

\subsection{Metallicity correction}

The measured value of $\mu_0$ must be corrected for the metallicity dependence
of the Cepheid P-L relation, using the following equation:

\vskip -6pt
\begin{equation}
\delta\mu_Z=\gamma_{VI}\ \big ( [O/H]_{N2841} - [O/H]_{LMC} \big ),
\end{equation}

\noindent{where $\gamma_{VI}$ is the metallicity dependence index for the V
and I filters, and $[O/H]_{gal} = \log (O/H)_{gal} / \log (O/H)_\odot$.  We
adopt $\gamma_{VI}=0.2\pm0.2$~mag/dex \citep{fre01}, which represents the
range of measurements found in the literature \citep{ko97,sa97, ken98}. The
adopted $[O/H]$ value for the LMC is $-0.4$~dex \citep{ken98}.}

The $[O/H]$ value for our field in \n2841 can be calculated using $([O\, II]
\lambda\lambda 3727)/H\beta$ and $([O\,III] \lambda\lambda 4959,5007)/H\beta$
line flux ratios listed in Table 3 of \citet{bre99} for four HII regions
located at galactocentric radii similar to those spanned by our Cepheids. The
sum of these two line flux ratios (known as $R_{23}$) can be turned into a
value of $[O/H]$ by using the appropriate Equation from \citet{zkh94}.  We
find $[O/H]_{N2841}= 0.4\pm0.15$~dex, which reflects the scatter in the
abundances of the four HII regions ($\pm 0.07$) and the uncertainty in the
calibration of the metallicity-$R_{23}$ relation for high abundances.

In conclusion, Equation (6) results in an overall correction of $\delta\mu_Z =
+0.16\pm0.15$~mag to our distance modulus, making it $\mu_{0,Z}=\muoe$~mag.
This uncertainty includes several sources of error, which we list in the
following sub-section.

\subsection{Error Budget}
The error budget for our distance determination is listed in Table~7, and
explained in greater detail here. The internal error due to the dispersion in
individual values of $\mu_0$ is a small contribution to the total error, which
is heavily dominated by external sources of error. These are, in decreasing
order of importance: the metallicity dependence of the Cepheid P-L relation,
the distance to the LMC, the absolute calibration of WFPC2, contamination of
the Cepheid magnitudes due to unresolved blends, and the uncertainty in the
reddening law of \n2841.

\noindent{{\it i.~Metallicity correction}: As stated in \S5.3, the
uncertainty in the adopted value of the metallicity correction index
$\gamma_{VI}$ is $\pm 0.2$~mag/dex, and the uncertainty in the
difference in abundances between the LMC and \n2841 is $\pm 0.15$~dex. The
combination of these two uncertainties results in a contribution to the error
budget of $\pm0.15$~mag.}

\noindent{{\it ii.~LMC Distance}: The uncertainty in the distance to the LMC
has been recently summarized by \citet{fre01}. We have adopted $\mu_0
{\rm(LMC)} = 18.50\pm0.10$~mag in order to maintain consistency with the
Cepheid distances presented in that paper.  However, different distance
indicators yield widely different (and non-overlapping) estimates of the LMC
distance modulus, ranging from 18.3 to 18.65~mag \citep[see Table 13
of][]{fre01}.}

\noindent{{\it iii.~Blending}: An additional source of uncertainty is the
possible contamination of our Cepheid magnitudes by nearby stars, physically
unrelated to the variables, but located at distances smaller than the
resolution provided by the telescope and detector. \citet{mo01} have studied
this contamination by comparing ground-based and HST images of long-period
($P>10d$) Cepheids in M33. They conclude that the median value of this effect
on HST data for galaxies at 10--15~Mpc is $\sim 7\%$ in V and $\sim 12\%$ in
I. If these numbers were to apply to our Cepheids, our $W$ magnitudes would be
biased by $5\%$, or $0.1$~mag. We take this value to represent a $1\sigma$
uncertainty due to the possible effects of blending.}

\noindent{{\it iv.~WFPC2 calibration}: \citet{fre01} have estimated the size
of the uncertainty related to the absolute photometric calibration of WFPC2 to
be $\pm0.07$~mag; this includes the uncertainties in the characterization of
the zeropoints, the color transformations, the camera gain ratios, and the
effects of CTE.}

\noindent{{\it v.~Reddening law}: The reddening-free Wesenheit magnitudes used
in Equation (5) were calculated using an assumed value of total-to-selective
extinction $R \equiv A_V/\ebv = 3.1 \equiv A_V/\evi = 2.45$. If the
interstellar medium of \n2841 were to follow a different extinction law than
the ISM of the Milky Way, the adoption of the standard value of $R$ would
introduce an additional systematic error to our distance estimate.  This issue
was addressed in a recent study by \citet{ma01}, who obtained NICMOS $H$ band
photometry of extragalactic Cepheids in a dozen galaxies.  They found that the
extinction corrections determined from either V and I data or from V and
$H$ data, using the standard extinction law, are consistent with each
other. Thus, the assumption of a standard reddening law is a valid one, and
does not introduce a significant source of error.}

The aforementioned sources of error (i-iv) can be combined in quadrature to
arrive at a total external uncertainty in our metallicity-corrected true
distance modulus of $+0.22, -0.19$~mag. The addition of the internal error
term ($\pm 0.06$~mag) increases the total uncertainty to $+0.23, -0.20$
~mag, which we simplify as $\pm0.23$~mag.

\section{Discussion}

\n2841 is one of eight galaxies with Cepheid distances associated with the Leo
Cloud \& Spur, out of a total of 65 members listed in \citet{NBG}. Table~8
lists some properties of those eight galaxies. All distances (except for the
one derived in this paper) are from \citet{fre01}. The mean velocity and
dispersion of the previously observed galaxies is $727\pm96$~km/s, while the
average distance (excluding the nearby NGC$\,$3621) is $11.5\pm0.6$~Mpc.  The
small velocity dispersion for this relatively spread out structure is somewhat
surprising, but consistent with bulk motion of the Leo Cloud.  Kinematically,
at a redshift of $638\pm3$~km/s, \n2841 is clearly a member of the Leo Spur,
and our derived distance of $\dist$~Mpc places it on the far side of this
structure.

Figure 10 shows the BVRI surface brightness profiles of \n2841 calculated from
CCD images obtained at the Fred L. Whipple Observatory 1.2-m telescope.  The
procedure we followed to obtain these profiles is identical to the one
described in \citet{ma00}. We find a $B=25\ {\rm mag}/\sq \arcsec$ isophotal
radius of $\sim 280\arcsec$, which corresponds to a linear isophotal radius of
$\sim 40$~kpc, given the distance of $\dist$~Mpc derived in the previous
section. Thus, \n2841 is an impressively large disk galaxy, about twice the
size of M31, a galaxy which shares many of its properties. Furthermore, the
values of extinction-corrected effective radius, $r_e^i \sim 6$~kpc, and
extinction-corrected effective surface brightness, $\langle\mu\rangle_e^i \sim
19.5\ {\rm I\ mag}/\sq \arcsec$, place \n2841 in a relatively unique position
in the surface brightness -- radius plane \citep[see Figure 2 of][]{djl00}.

\section{Conclusions}
We have discovered a total of $\nc$ Cepheids and $\nm$ other variables in
the Sb spiral galaxy \n2841, a member of the Leo Spur. We have applied fiducial
OGLE LMC V- and I-band Cepheid Period-Luminosity relations to a sub-sample
of $\ng$ Cepheids. We adopted an LMC distance of $\mu_0$(LMC)$=18.50\pm0.10$
~mag, and a metallicity dependence of $\gamma_{VI}=0.2\pm0.2$ ~mag/dex. Based
on these assumptions, we obtain a distance modulus of $\muoe$~mag,
corresponding to a distance of $\dist$~Mpc. Our distance estimate is consistent
with an association of this galaxy to the Leo Spur. The mean reddening of the
Cepheid sample is $\evi=\mvi$~mag.

A companion paper by \citet{bo01} will address the impact of the Cepheid
distance to \n2841 on the absolute calibration of the Tully-Fisher relation, as
well as the implication of our distance for the MOND theory. A second companion
paper by \citet{ga01} will combine our Cepheid distance with observations of
\sn1999by to further the absolute calibration of the type Ia SNe distance
indicator.

We would like to thank the referee, John Graham, for his helpful
comments, and the STScI and NASA support staff that have made these
observations possible. This work has made use of the NASA/IPAC
Extragalactic Database (NED), which is operated by the Jet Propulsion
Laboratory, Caltech, under contract with the National Aeronautics and
Space Administration.

\clearpage

\setcounter{figure}{0}

\begin{figure}
\plotone{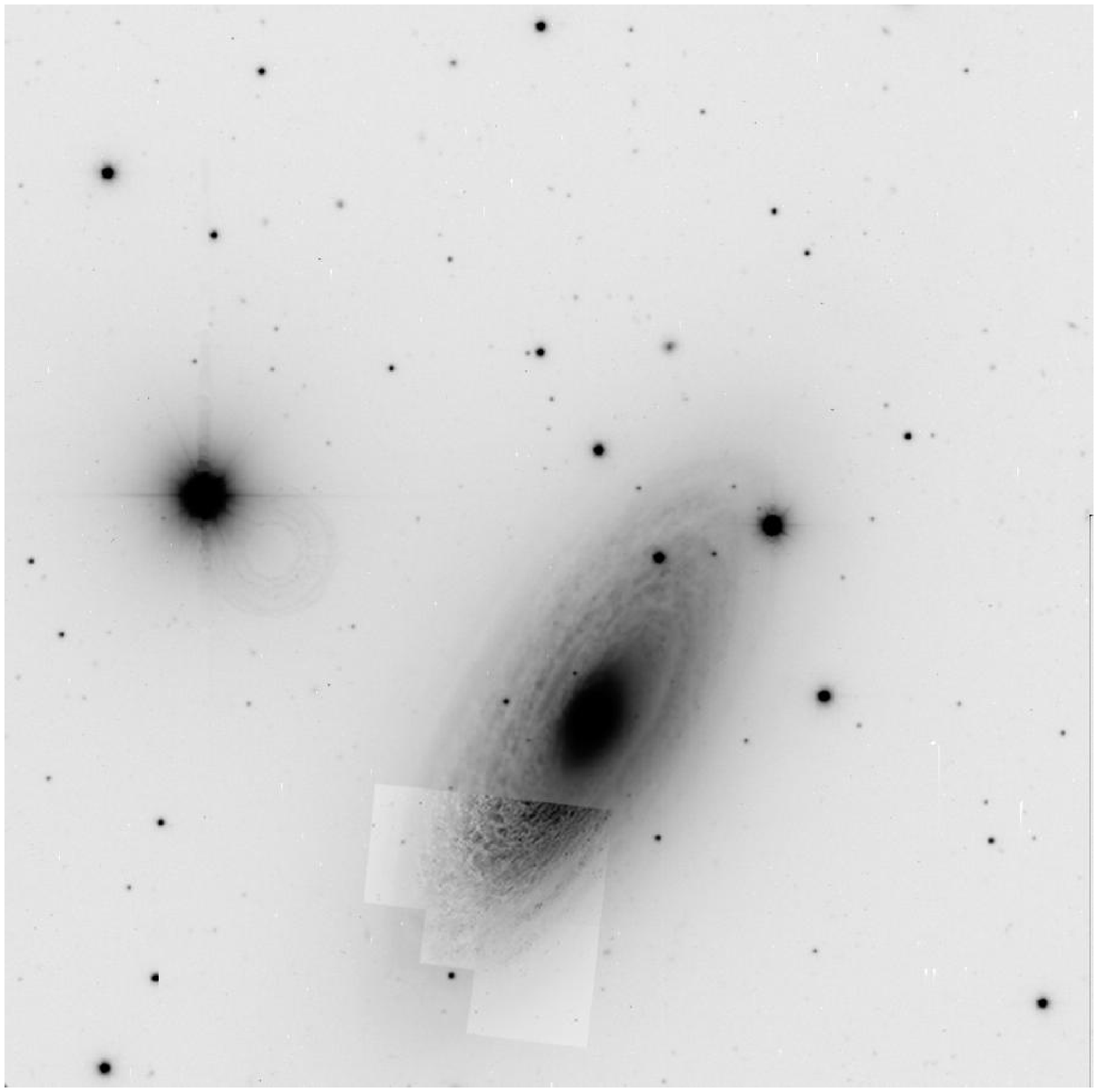}
\caption{Ground-based image of \n2841, obtained at the FLWO 1.2-m
telescope, with a super-imposed mosaic of the WFPC2 field of view (see
Figure~2). North is up and east is to the left.}
\end{figure}

\clearpage

\begin{figure}
\plotone{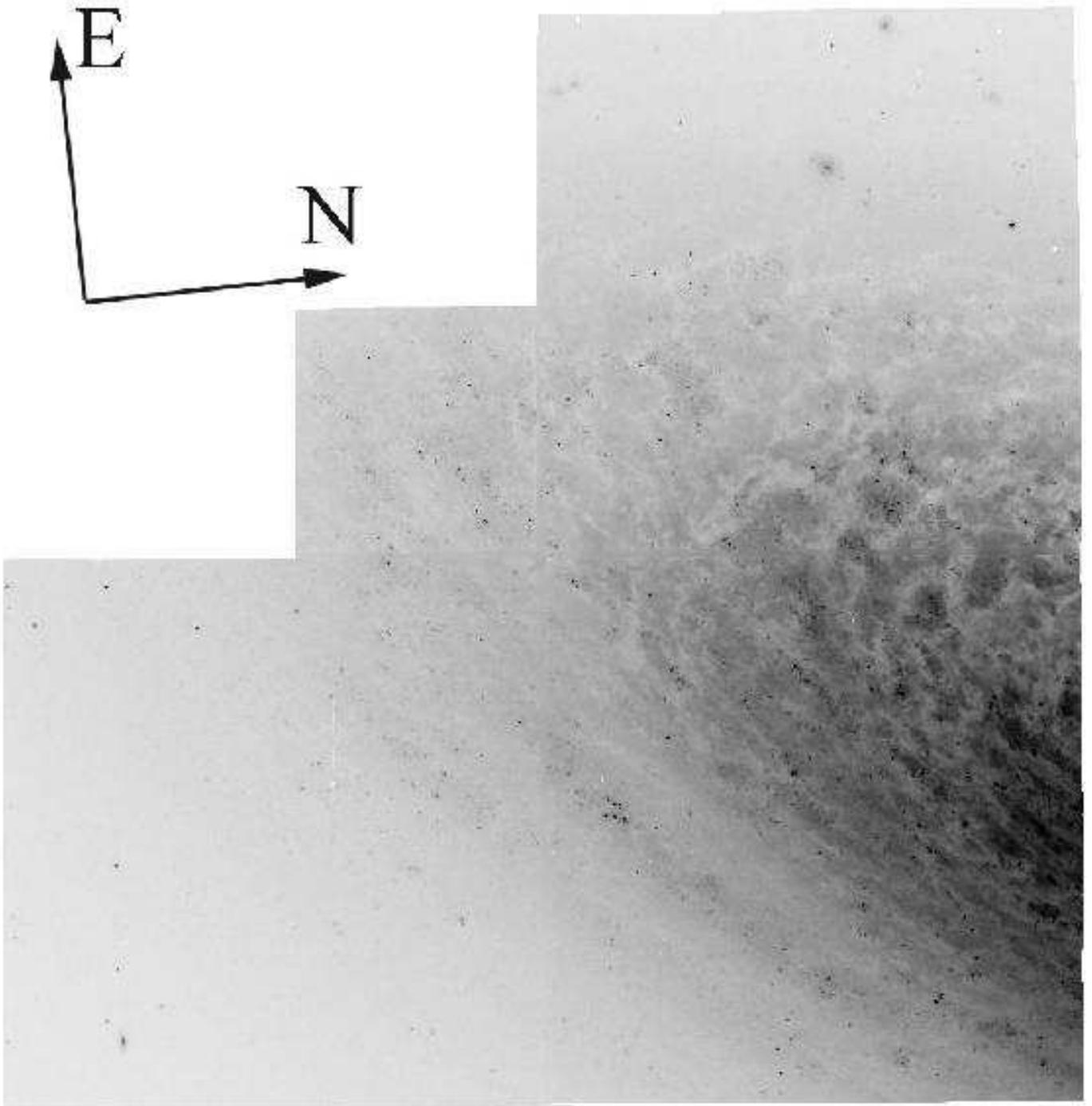}
\caption{Mosaic of the WFPC2 field of view of \n2841. Individual images of each
chip, showing the location of the variables, can be found in Figures~3a-d.}
\end{figure}

\clearpage

\begin{figure}
\plotone{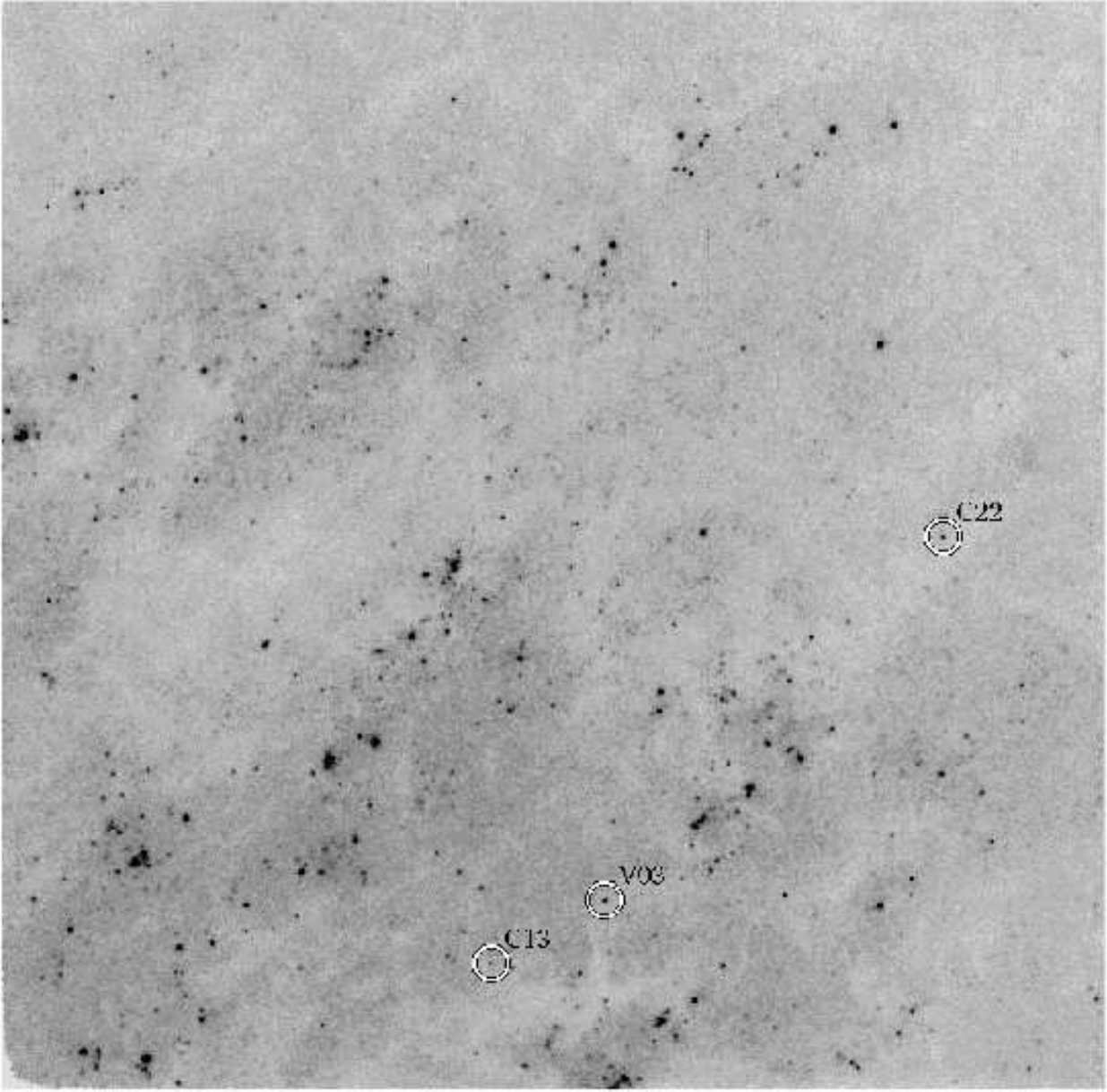}
\caption{(a) Medianed images of the four WFPC2 chips. The circles indicate the
position of each of the variables, labeled as in Table~3. Each of the images
is oriented such that the bottom left corner has the pixel coordinates (0,0) in
each image.}
\end{figure}

\clearpage

\setcounter{figure}{2}
\begin{figure}
\plotone{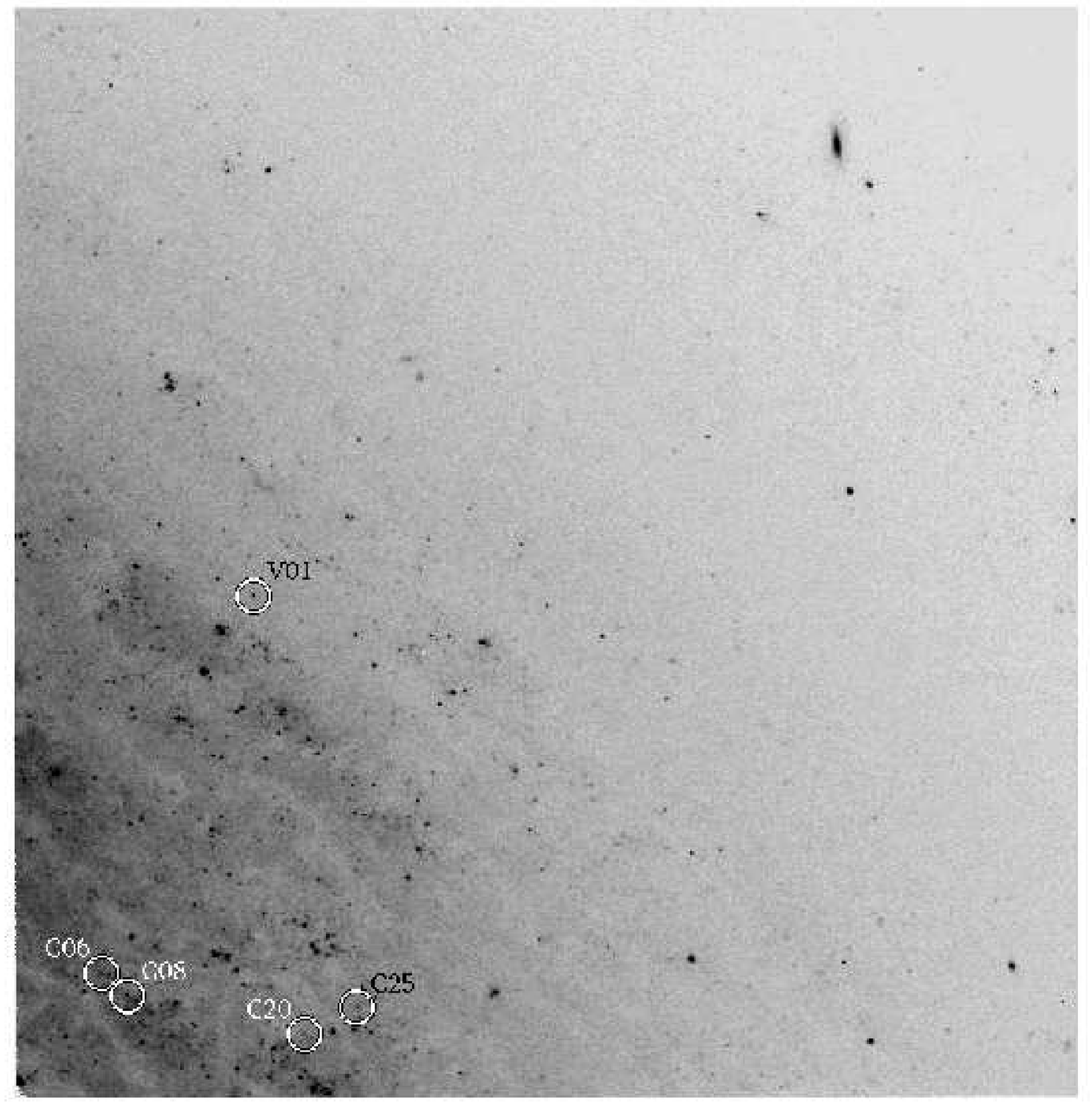}
\caption{(b) continued.}
\end{figure}

\clearpage

\setcounter{figure}{2}
\begin{figure}
\plotone{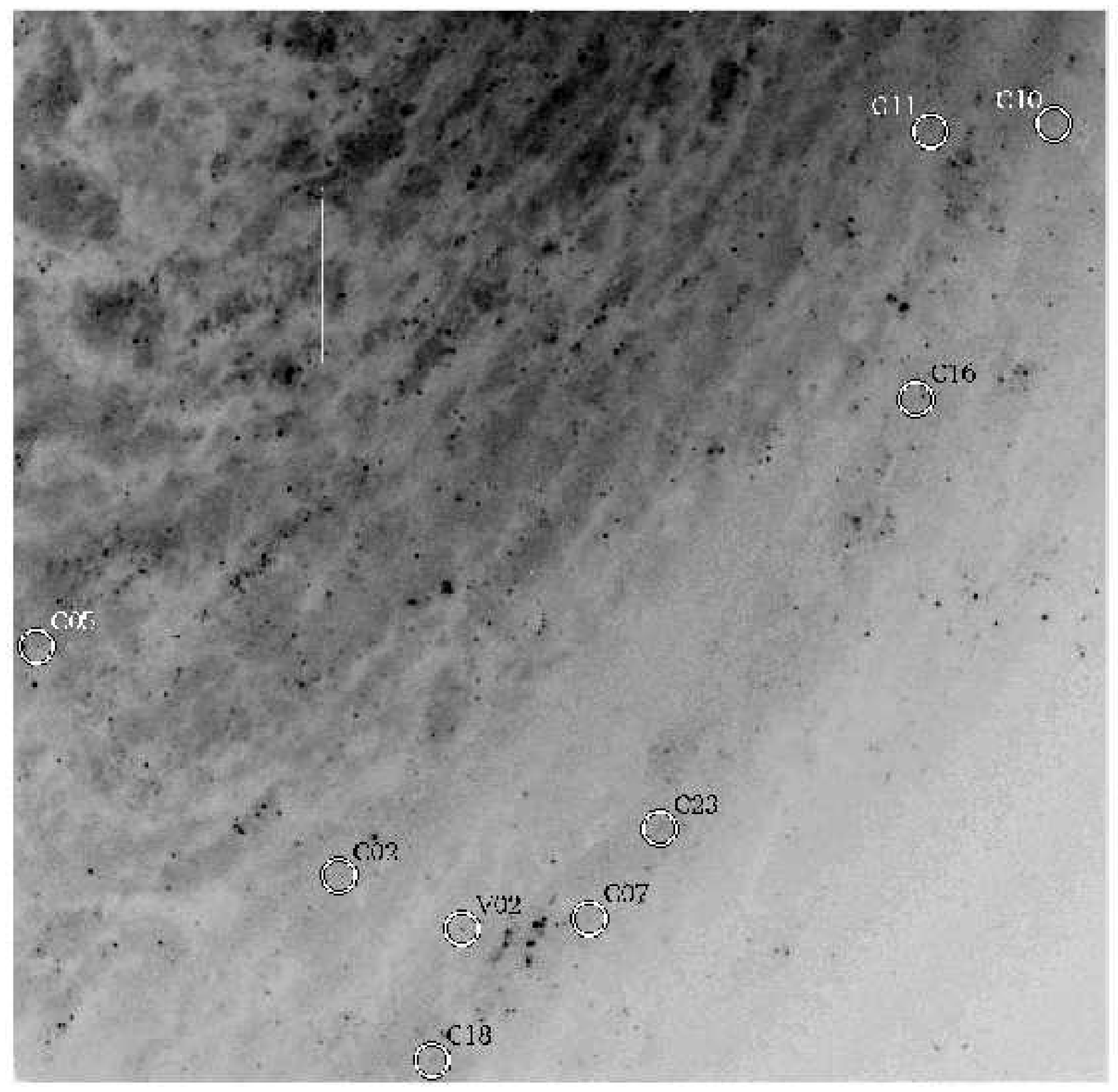}
\caption{(c) continued.}
\end{figure}

\clearpage

\setcounter{figure}{2}
\begin{figure}
\plotone{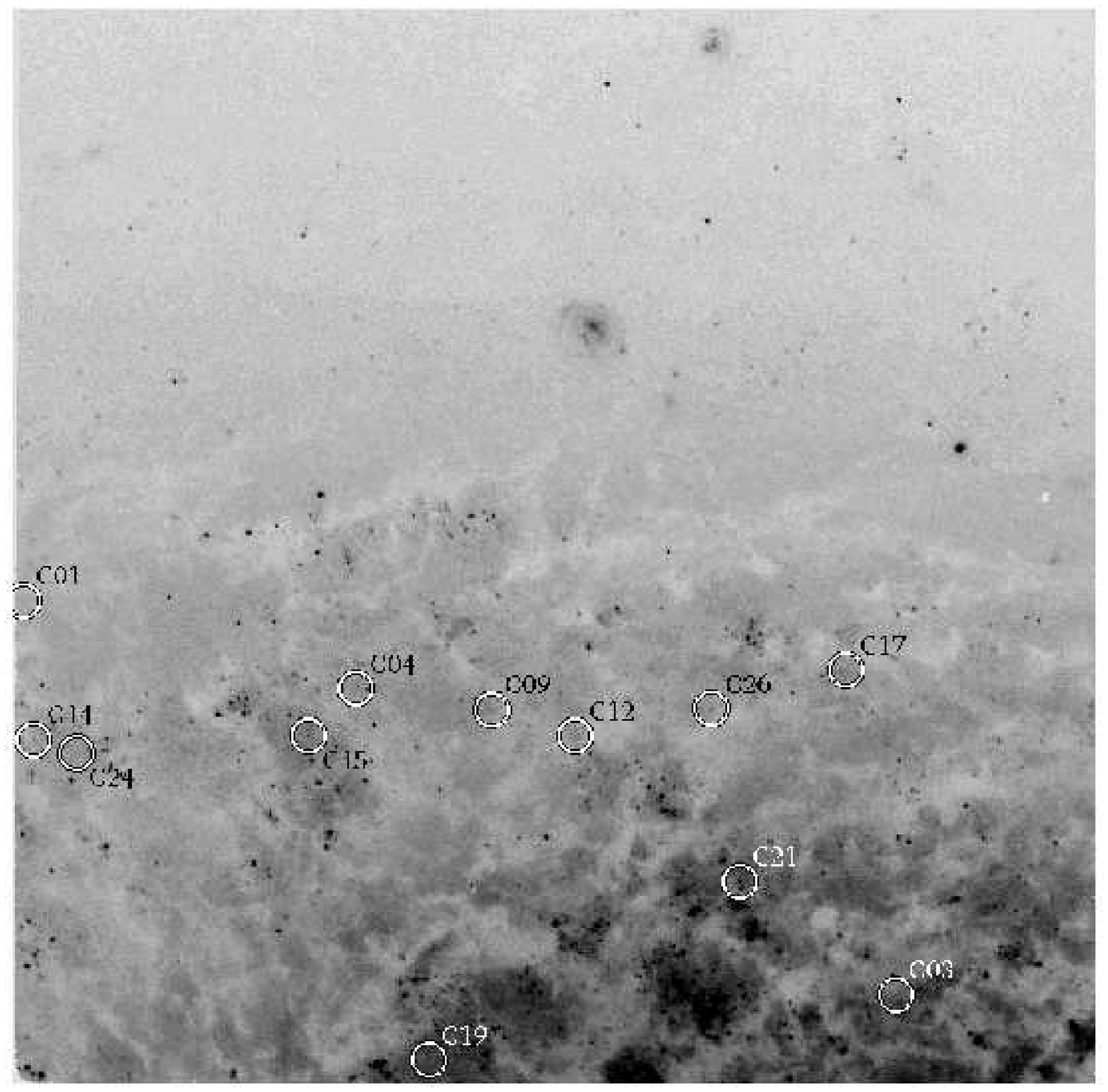}
\caption{(d) continued.}
\end{figure}

\clearpage

\begin{figure}
\plotone{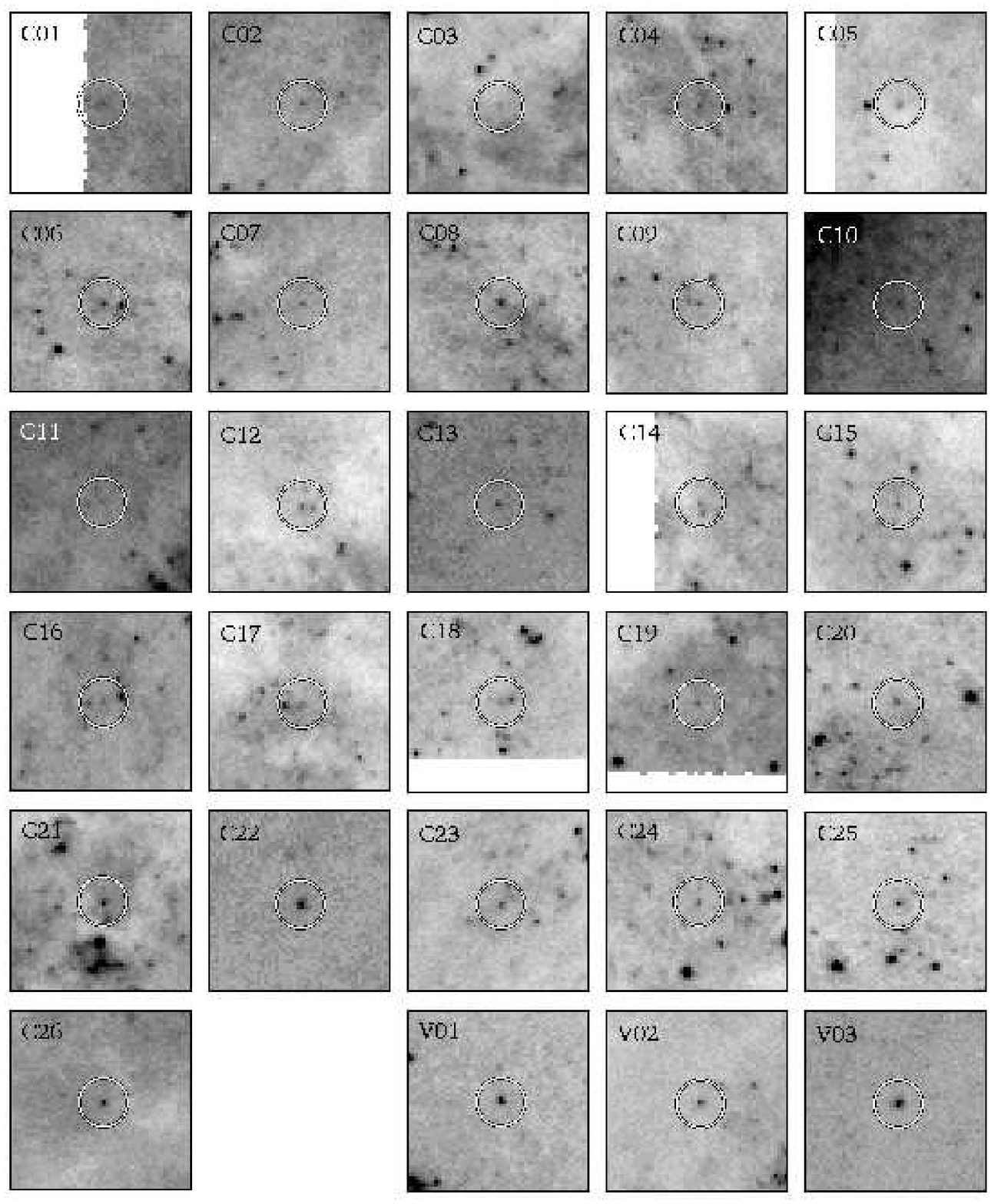}
\caption{Individual finding charts for the variables. The position of each    
object is shown by the circle. The charts are 50 pixels on a side, i.e.,
$2\Sec15\times 2\Sec15$ for objects in the PC chip and $5\arcsec\times
5\arcsec$ for objects in the the WF chips.}
\end{figure}

\clearpage

\begin{figure}
\plotone{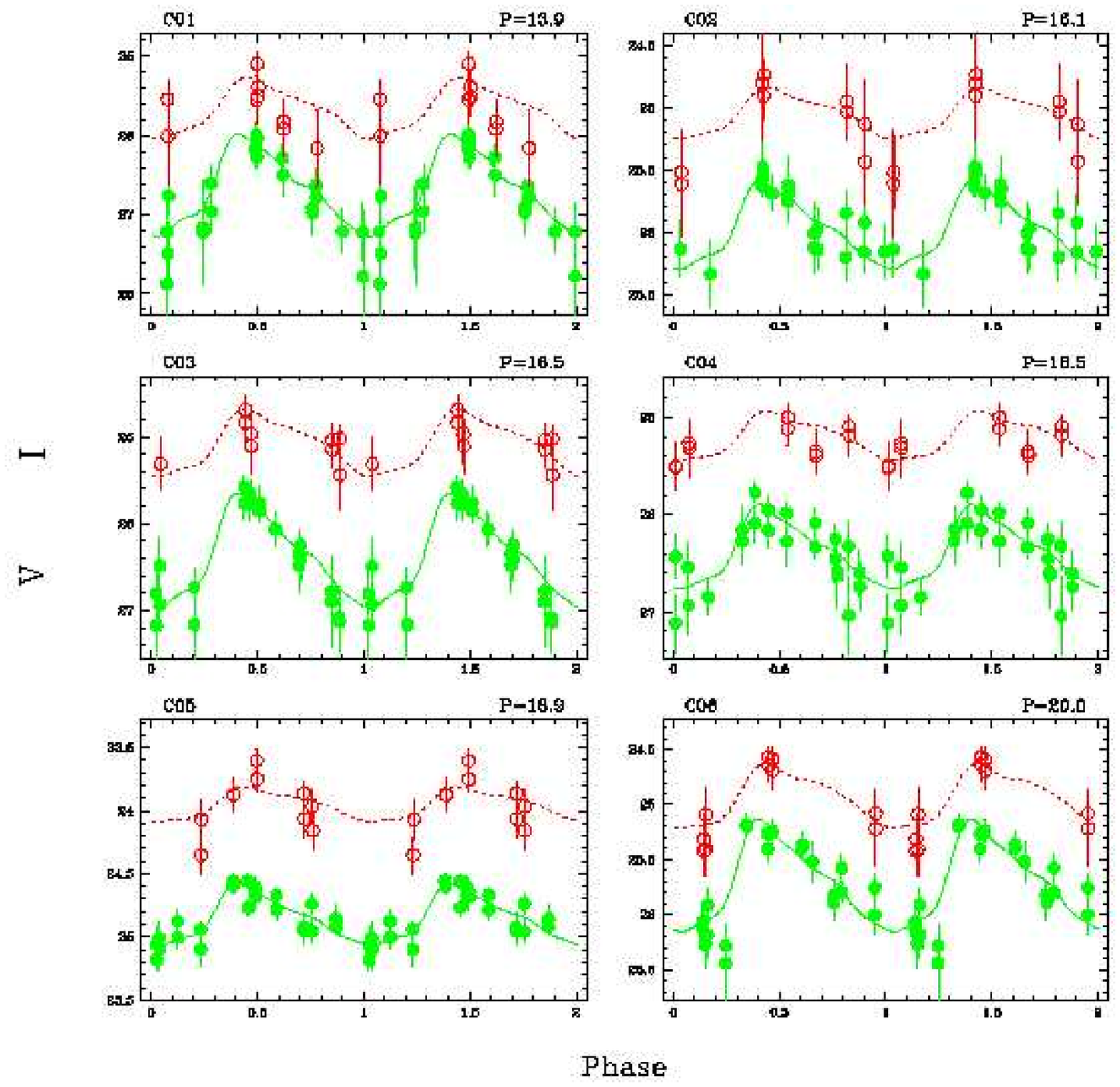}
\caption{(a) V and I band phase-magnitude plots for Cepheids C01-C06. Two
complete cycles are plotted to facilitate ease of interpretation. Filled
and open circles are used to represent the V and I magnitudes
listed in Table~5, respectively. The uncertainties associated with each
measurement, also listed in that Table, are plotted using error bars. The solid
and dashed lines overplotted on each panel indicate the best-fit V and I
band template light curves, respectively.}
\end{figure}

\clearpage

\setcounter{figure}{4}
\begin{figure}
\plotone{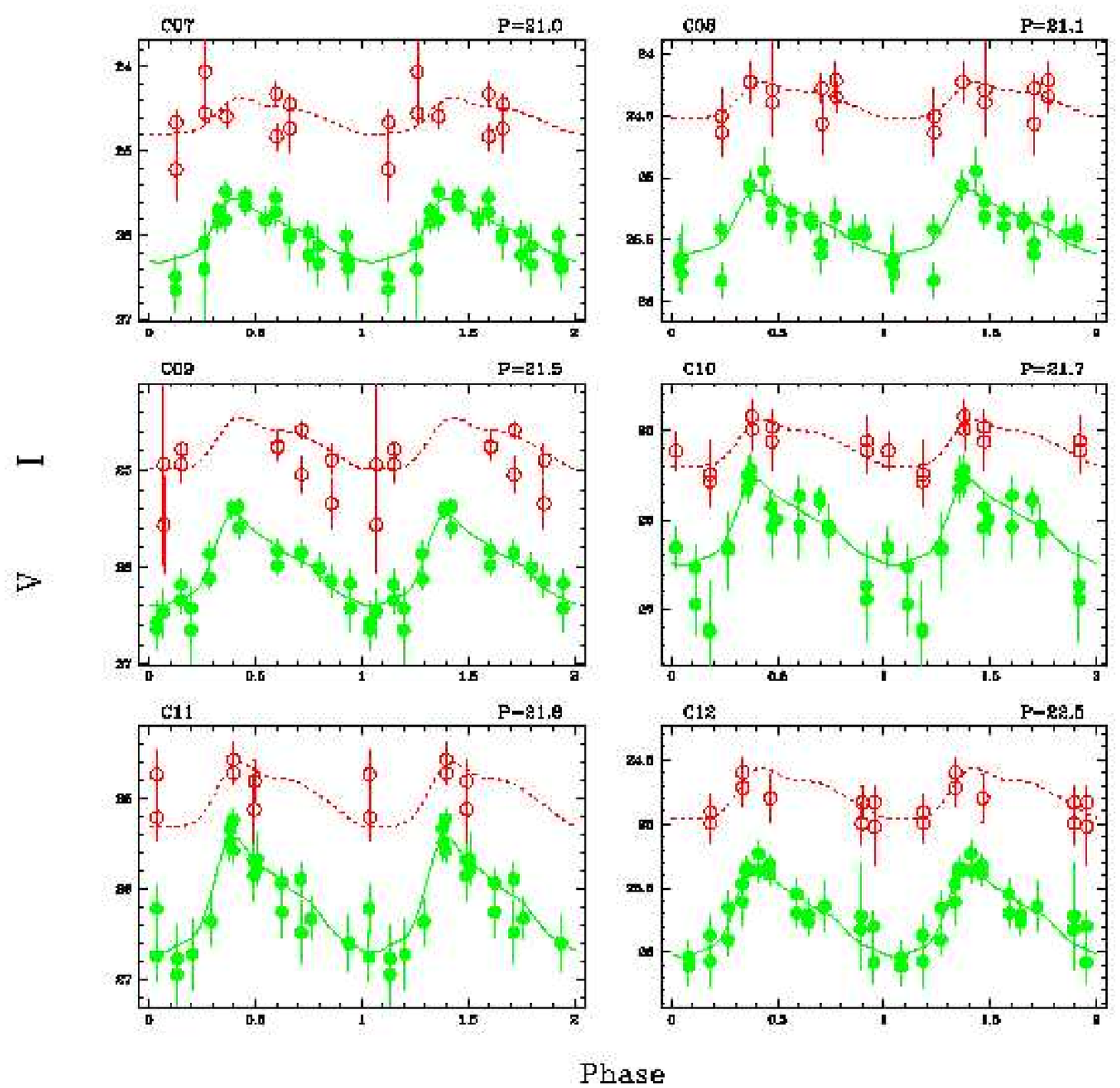}
\caption{(b) same as Figure~5a, for objects C07-C12.}
\end{figure}

\clearpage

\setcounter{figure}{4}
\begin{figure}
\plotone{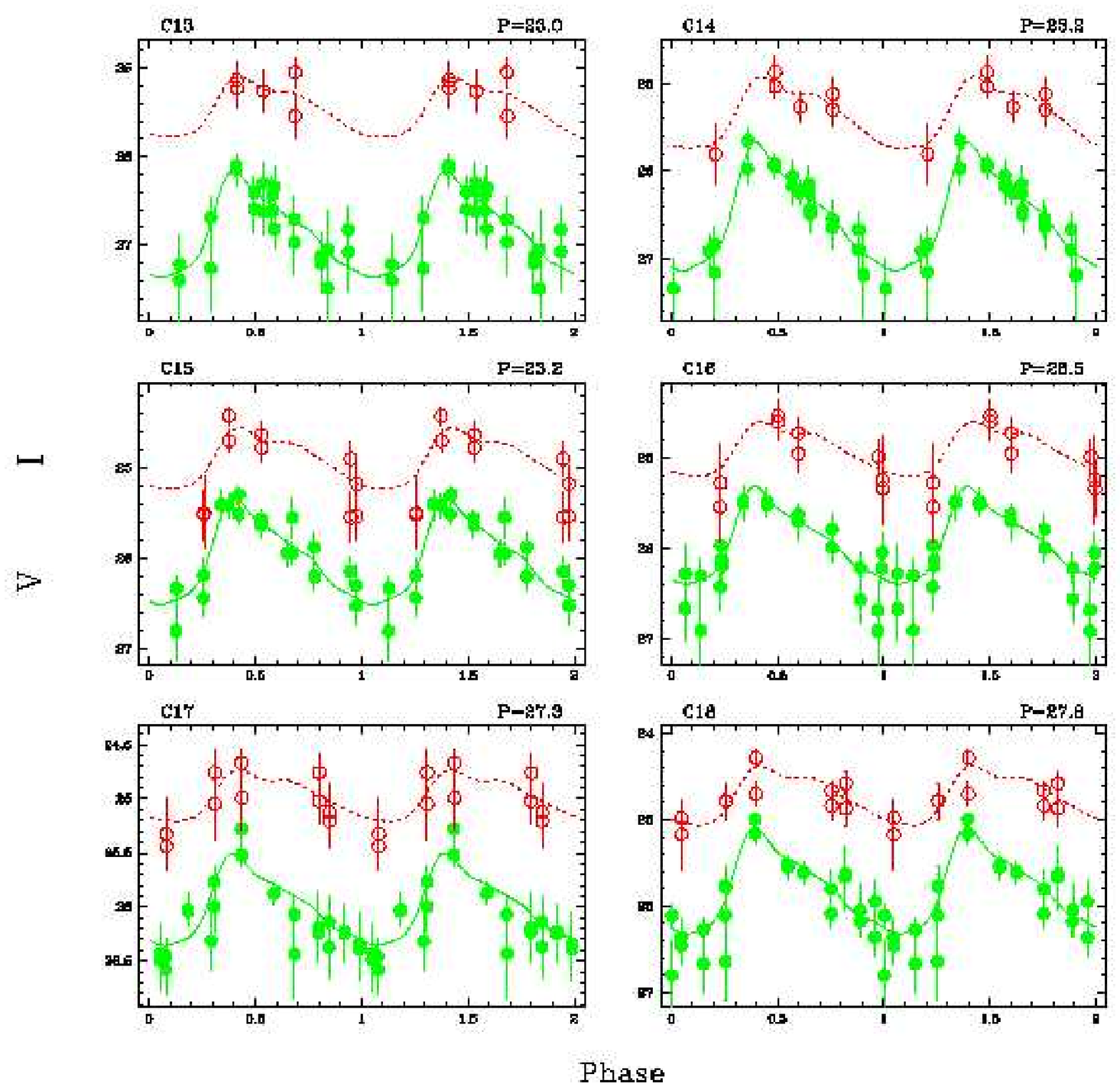}
\caption{(c) same as Figure~5a, for objects C13-C18.}
\end{figure}

\clearpage

\setcounter{figure}{4}
\begin{figure}
\plotone{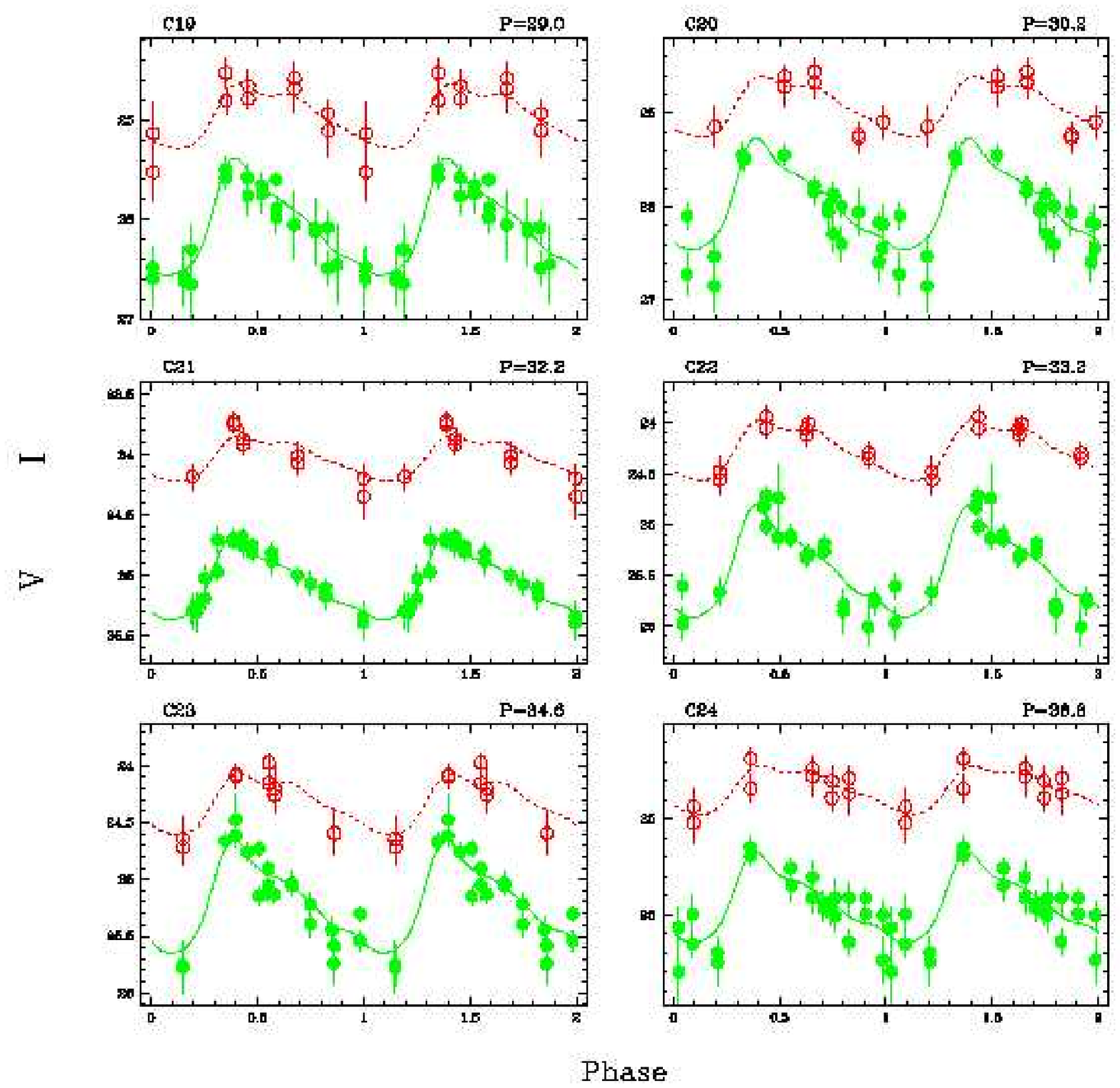}
\caption{(d) same as Figure~5a, for objects C19-C24.}
\end{figure}

\clearpage

\setcounter{figure}{4}
\begin{figure}
\plotone{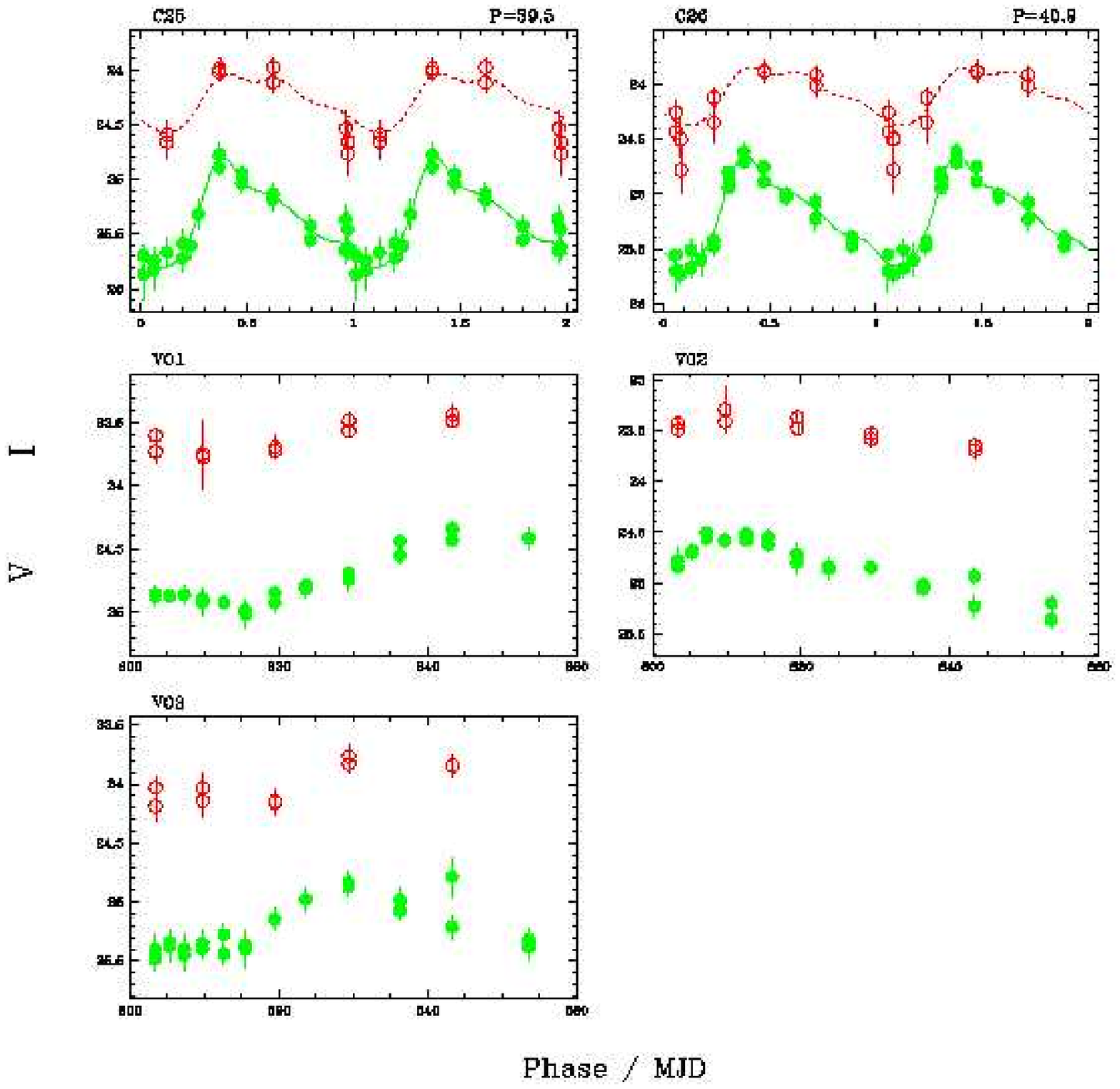}
\caption{(e) Top: same as Figure~5a, for objects C25-C26. Middle and bottom: V
and I band time series for the variables V01-V03, plotted as magnitude versus
modified Julian Date.}
\end{figure}

\clearpage

\begin{figure}
\plotone{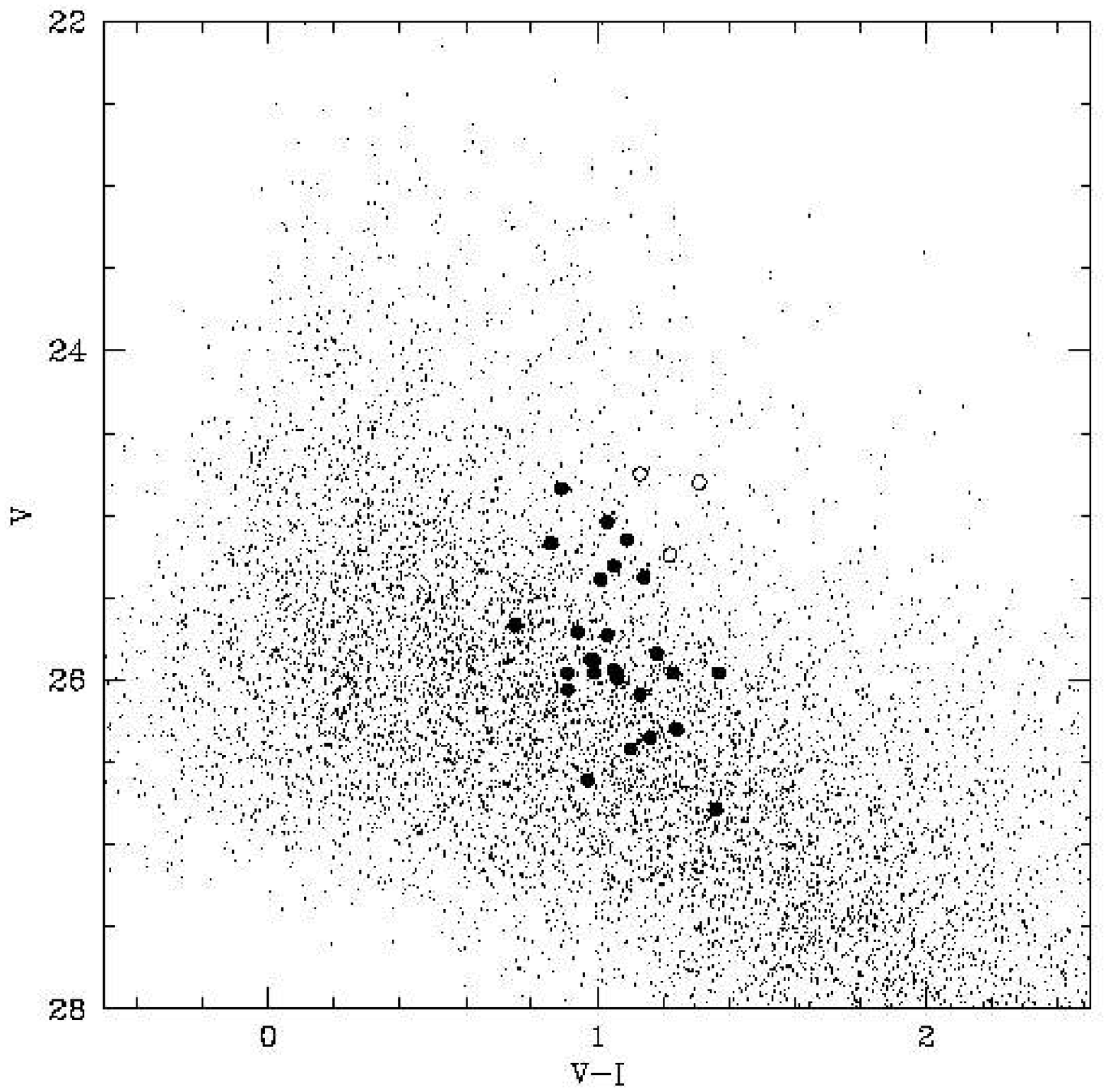}
\caption{Color-magnitude diagram of the $\sim 9000$ stars detected in our
\n2841 images. Filled and open circles are used to plot the Cepheids and the
other variables, respectively.}
\end{figure}

\clearpage

\begin{figure}
\plotone{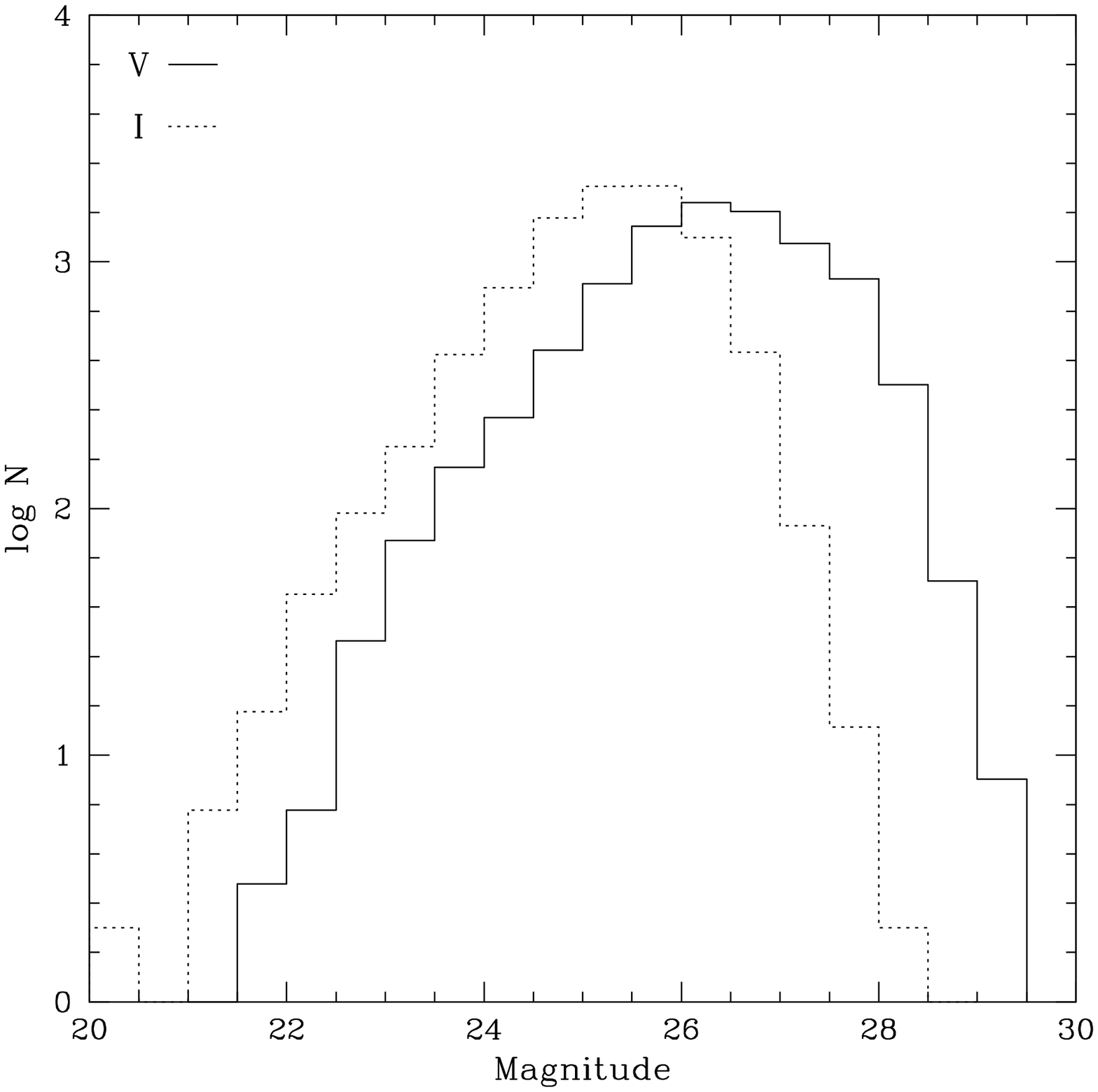}
\caption{Observed differential luminosity functions in the \n2841 WFPC2 field.
Solid and dashed lines are used to plot the V and I histograms, 
respectively. Based on these distributions, we determine completeness limits
of V$\sim 26.5$~mag and I$\sim 25.5$~mag.}
\end{figure}

\clearpage

\begin{figure}
\plotone{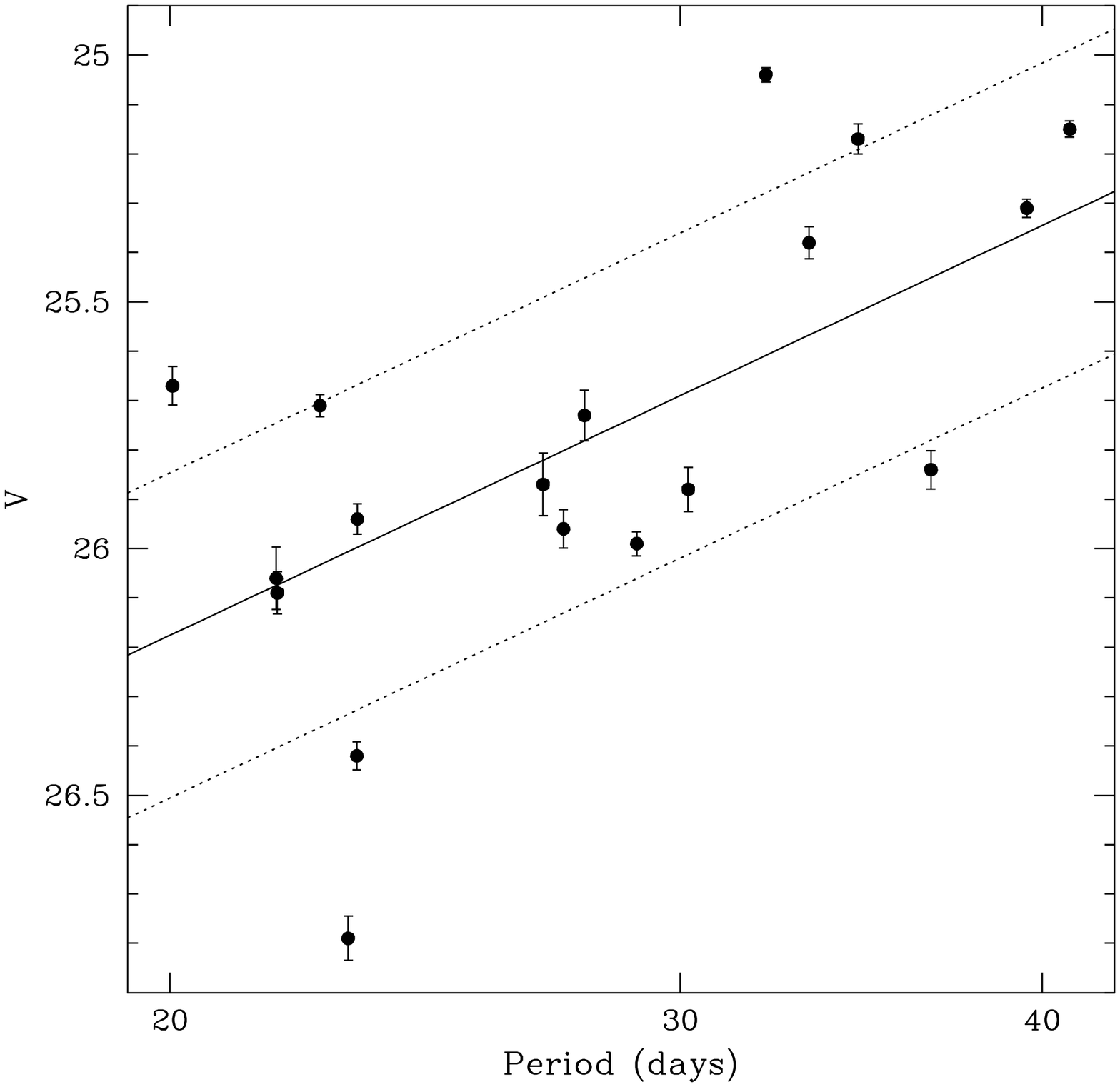}
\caption{V-band Period-Luminosity relation for the selected Cepheids in
our sample (see Table 4 and \S 5.2). The solid line is the best fit to the
fiducial LMC P-L relation listed in Equation (3) and the dotted lines
correspond to the {\it r.m.s.} dispersion of the fit. The apparent distance
modulus obtained from the fit is $\muv$~mag.}
\end{figure}

\clearpage

\begin{figure}
\plotone{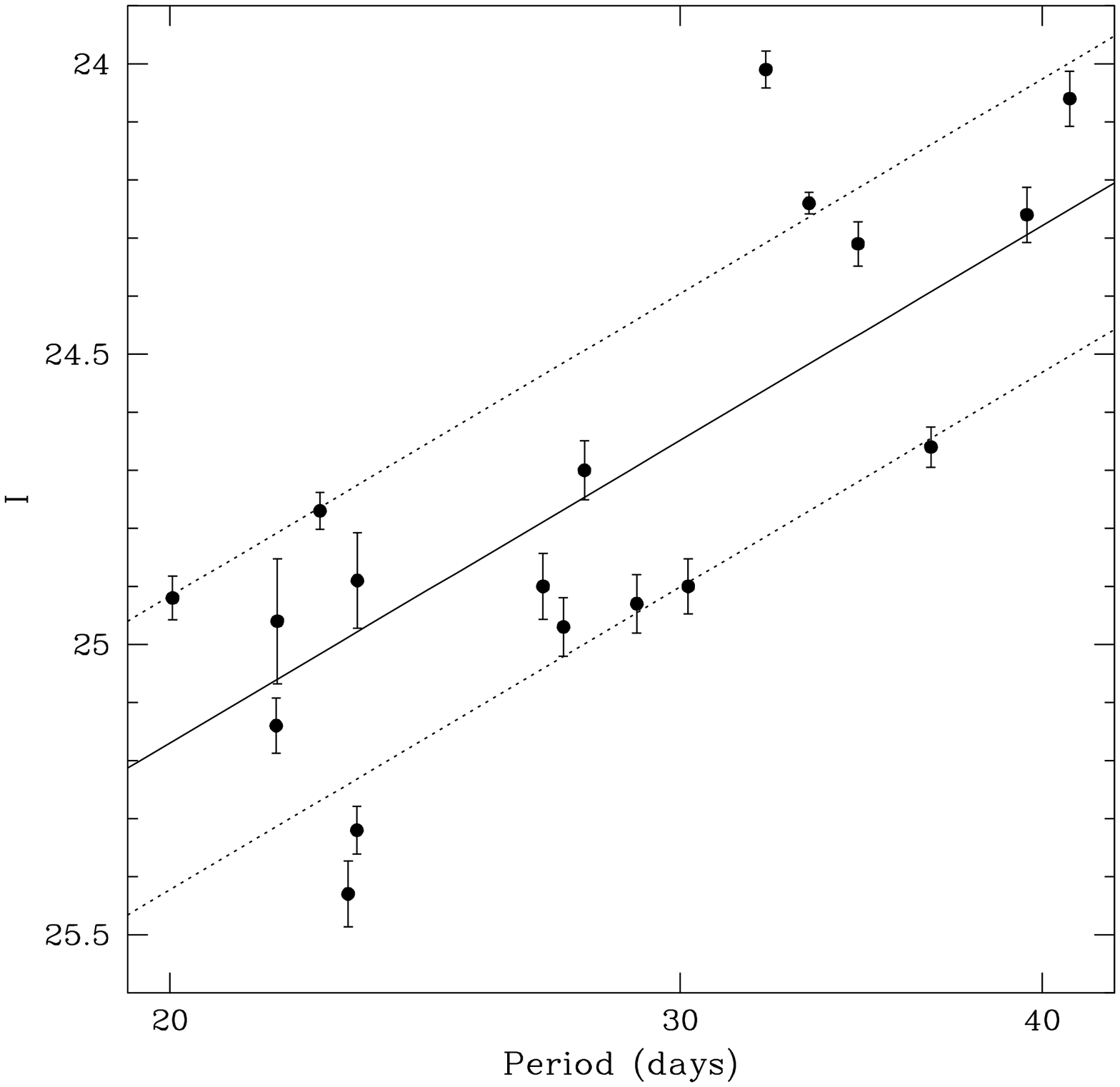}
\caption{ I-band Period Luminosity relation for the selected Cepheids in
our sample (see Table 4 and \S 5.2). The solid line is the best fit to the
fiducial LMC P-L relation listed in Equation (4) and the dotted lines
correspond to the {\it r.m.s.} dispersion of the fit. The apparent distance
modulus obtained from the fit is $\mui$~mag.}
\end{figure}

\clearpage

\begin{figure}
\plotone{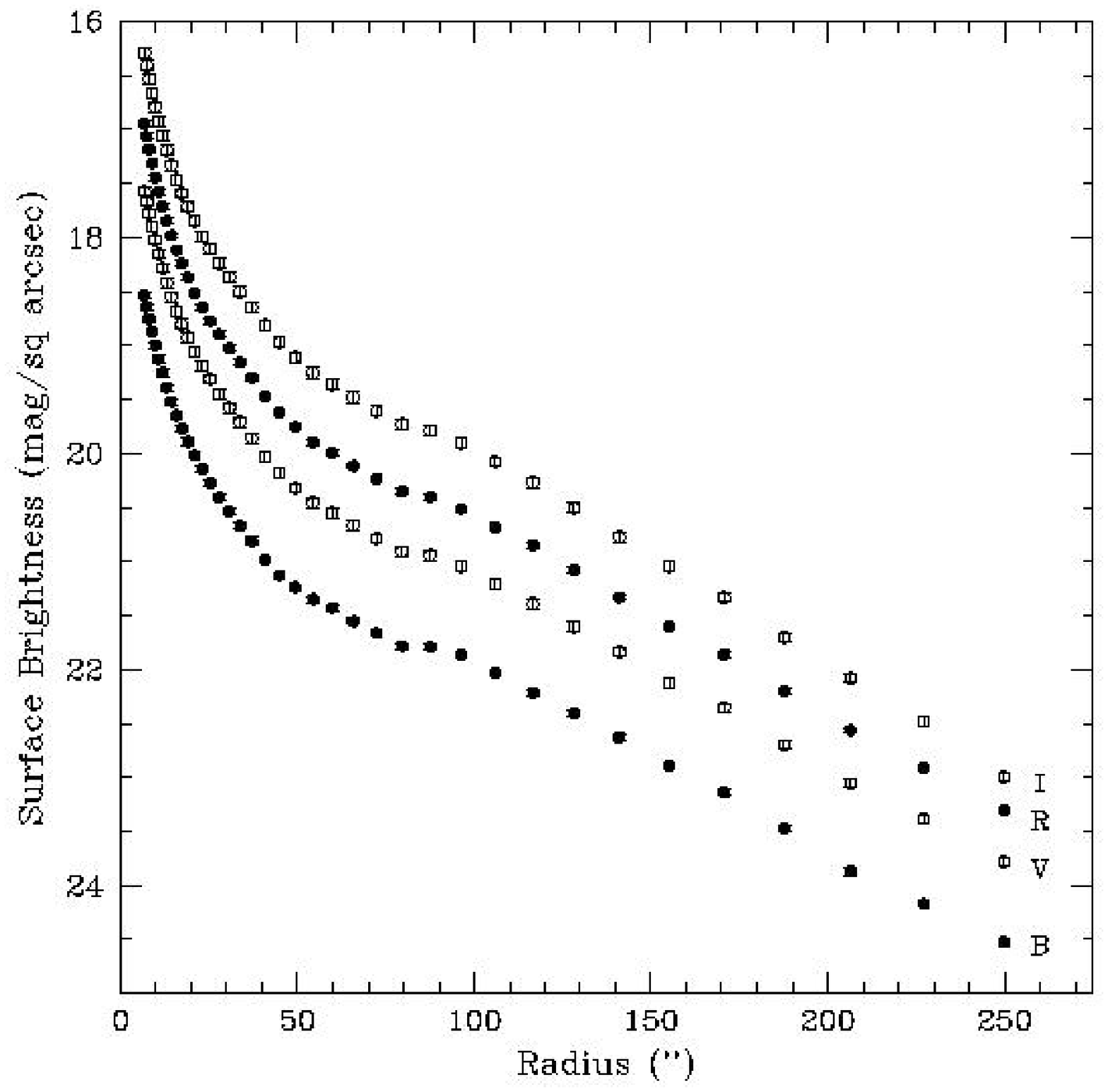}
\caption{BVRI surface brightness profiles of \n2841, calculated from images
obtained at the FLWO 1.2-m telescope and following the precepts of
\citet{ma00}.}
\end{figure}

\begin{deluxetable}{llll}
\tablecaption{HST Observations of \n2841}
\tablenum{1}
\tablewidth{0pt}
\tablehead{\colhead{Date} & \colhead{JD} &
\colhead{Exp. time} & \colhead{Filter}}
\startdata
2000 Feb 29 & 2451603.35 & $2\times 1100s$ & F555W \\
            & 2451603.41 & $2\times 1100s$ & F814W \\
2000 Mar 02 & 2451605.29 & $2\times 1100s$ & F555W \\
2000 Mar 04 & 2451607.24 & $2\times 1100s$ & F555W \\
2000 Mar 06 & 2451609.66 & $2\times 1100s$ & F555W \\
            & 2451609.72 & $2\times 1100s$ & F814W \\
2000 Mar 09 & 2451612.54 & $2\times 1100s$ & F555W \\
2000 Mar 12 & 2451615.49 & $2\times 1100s$ & F555W \\
2000 Mar 16 & 2451619.38 & $2\times 1100s$ & F555W \\
            & 2451619.44 & $2\times 1100s$ & F814W \\
2000 Mar 20 & 2451623.60 & $2\times 1100s$ & F555W \\
2000 Mar 26 & 2451629.30 & $2\times 1100s$ & F555W \\
            & 2451629.36 & $2\times 1100s$ & F814W \\
2000 Apr 02 & 2451636.27 & $2\times 1100s$ & F555W \\
2000 Apr 09 & 2451643.24 & $2\times 1100s$ & F555W \\
            & 2451643.31 & $2\times 1100s$ & F814W \\
2000 Apr 19 & 2451653.56 & $2\times 1100s$ & F555W \\
\enddata
\end{deluxetable}

\begin{deluxetable}{lll}
\tablecaption{WFPC2 Zeropoints}
\tablenum{2}
\tablewidth{0pt}
\tablehead{\colhead{Chip} & \multicolumn{2}{c}{Zeropoints} \\
\colhead{} & \colhead{F555W} & \colhead{F814W}}
\startdata
PC1 & $0.9510\pm0.0028$ & $1.9201\pm0.0017$ \\
WF2 & $0.9584\pm0.0016$ & $1.8478\pm0.0011$ \\
WF3 & $0.9513\pm0.0014$ & $1.8724\pm0.0010$ \\
WF4 & $0.9697\pm0.0015$ & $1.9007\pm0.0010$ \\
\enddata
\end{deluxetable}

\clearpage

\begin{deluxetable}{lcrrrrcc}
\tablecaption{Secondary standards in \n2841}
\tablenum{3}
\tablewidth{0pt}
\tablehead{\colhead{ID} & \colhead{Chip} & \colhead{x} & \colhead{y} & \multicolumn{2}{c}{R.A. (J2000.0) Dec.} & \colhead{V (mag)} & \colhead{I (mag)}}
\startdata
S01 & 1 & 132.2 & 467.7 & 09:22:10.649 & 50:56:12.66 & $25.40\pm0.02$ & $24.55\pm0.04$ \\ 
S02 & 1 & 141.4 & 530.3 & 09:22:10.725 & 50:56:09.88 & $25.26\pm0.02$ & $25.17\pm0.07$ \\ 
S03 & 1 & 238.3 & 600.8 & 09:22:11.223 & 50:56:07.16 & $26.47\pm0.04$ & $24.00\pm0.02$ \\ 
S04 & 1 & 540.2 & 144.5 & 09:22:12.433 & 50:56:29.20 & $25.18\pm0.02$ & $25.14\pm0.10$ \\ 
S05 & 1 & 508.3 & 685.9 & 09:22:12.556 & 50:56:04.64 & $25.22\pm0.02$ & $25.23\pm0.07$ \\ 
S06 & 1 & 591.2 & 687.4 & 09:22:12.952 & 50:56:04.99 & $25.84\pm0.04$ & $24.29\pm0.03$ \\ 
S07 & 1 & 659.5 &  98.5 & 09:22:12.976 & 50:56:31.81 & $25.27\pm0.02$ & $25.30\pm0.07$ \\ 
S08 & 1 & 715.8 & 262.0 & 09:22:13.329 & 50:56:24.75 & $24.76\pm0.01$ & $24.54\pm0.06$ \\ 
\hline
S09 & 2 & 117.0 & 744.0 & 09:22:02.603 & 50:56:16.07 & $24.92\pm0.02$ & $23.20\pm0.02$ \\ 
S10 & 2 & 390.5 & 543.7 & 09:22:05.010 & 50:55:51.50 & $25.28\pm0.02$ & $23.26\pm0.01$ \\ 
S11 & 2 & 272.8 & 410.5 & 09:22:06.260 & 50:56:04.66 & $24.65\pm0.01$ & $23.92\pm0.02$ \\ 
S12 & 2 & 193.1 & 396.0 & 09:22:06.315 & 50:56:12.69 & $24.22\pm0.01$ & $24.06\pm0.02$ \\ 
S13 & 2 &  68.9 & 371.7 & 09:22:06.418 & 50:56:25.17 & $24.78\pm0.01$ & $24.07\pm0.03$ \\ 
S14 & 2 & 464.0 & 355.9 & 09:22:07.063 & 50:55:46.39 & $24.52\pm0.01$ & $24.66\pm0.03$ \\ 
S15 & 2 & 349.2 & 308.1 & 09:22:07.424 & 50:55:58.29 & $24.65\pm0.01$ & $24.66\pm0.05$ \\ 
S16 & 2 & 189.2 & 271.2 & 09:22:07.615 & 50:56:14.49 & $24.86\pm0.01$ & $24.30\pm0.03$ \\ 
S17 & 2 & 208.3 & 181.5 & 09:22:08.574 & 50:56:13.62 & $24.86\pm0.02$ & $24.20\pm0.04$ \\ 
S18 & 2 & 225.7 & 160.3 & 09:22:08.816 & 50:56:12.14 & $24.50\pm0.01$ & $23.69\pm0.02$ \\ 
S19 & 2 & 332.0 &  64.2 & 09:22:09.944 & 50:56:02.75 & $24.69\pm0.02$ & $24.14\pm0.03$ \\ 
\hline
S20 & 3 & 730.9 & 495.3 & 09:22:02.202 & 50:57:07.63 & $24.47\pm0.01$ & $24.31\pm0.04$ \\ 
S21 & 3 & 723.7 & 457.7 & 09:22:02.319 & 50:57:04.01 & $26.69\pm0.06$ & $24.08\pm0.03$ \\ 
S22 & 3 & 705.2 & 400.4 & 09:22:02.576 & 50:56:58.57 & $25.38\pm0.03$ & $25.10\pm0.09$ \\ 
S23 & 3 & 716.0 & 214.2 & 09:22:02.682 & 50:56:40.13 & $25.19\pm0.02$ & $24.92\pm0.07$ \\ 
S24 & 3 & 689.1 & 373.8 & 09:22:02.774 & 50:56:56.12 & $26.39\pm0.07$ & $24.26\pm0.04$ \\ 
S25 & 3 & 702.8 & 156.2 & 09:22:02.888 & 50:56:34.59 & $26.83\pm0.07$ & $24.14\pm0.03$ \\ 
S26 & 3 & 631.8 & 364.0 & 09:22:03.382 & 50:56:55.77 & $23.78\pm0.01$ & $22.80\pm0.02$ \\ 
S27 & 3 & 634.1 & 284.6 & 09:22:03.450 & 50:56:47.92 & $24.44\pm0.01$ & $24.11\pm0.03$ \\ 
S28 & 3 & 607.5 & 139.7 & 09:22:03.897 & 50:56:33.96 & $25.09\pm0.01$ & $22.40\pm0.02$ \\ 
S29 & 3 & 509.6 & 207.4 & 09:22:04.840 & 50:56:41.64 & $26.46\pm0.06$ & $24.30\pm0.03$ \\ 
S30 & 3 & 473.9 & 177.8 & 09:22:05.247 & 50:56:39.12 & $23.22\pm0.01$ & $22.34\pm0.01$ \\ 
S31 & 3 & 462.7 & 101.3 & 09:22:05.452 & 50:56:31.73 & $24.77\pm0.02$ & $24.83\pm0.06$ \\ 
S32 & 3 & 434.2 & 248.7 & 09:22:05.580 & 50:56:46.53 & $27.02\pm0.07$ & $23.91\pm0.02$ \\ 
S33 & 3 & 425.0 & 225.3 & 09:22:05.704 & 50:56:44.32 & $26.31\pm0.04$ & $24.42\pm0.04$ \\ 
S34 & 3 & 385.3 & 100.9 & 09:22:06.261 & 50:56:32.52 & $24.48\pm0.01$ & $24.24\pm0.03$ \\ 
S35 & 3 & 360.1 &  82.5 & 09:22:06.545 & 50:56:31.00 & $26.07\pm0.03$ & $23.60\pm0.02$ \\ 
S36 & 3 & 282.6 & 110.9 & 09:22:07.322 & 50:56:34.63 & $26.55\pm0.06$ & $23.91\pm0.02$ \\ 
S37 & 3 & 280.5 &  93.6 & 09:22:07.363 & 50:56:32.96 & $25.19\pm0.02$ & $24.77\pm0.05$ \\ 
S38 & 3 & 249.9 &  78.8 & 09:22:07.698 & 50:56:31.85 & $25.15\pm0.02$ & $24.64\pm0.05$ \\ 
S39 & 3 & 220.4 & 141.7 & 09:22:07.936 & 50:56:38.34 & $24.75\pm0.01$ & $24.45\pm0.04$ \\ 
S40 & 3 & 163.9 & 162.4 & 09:22:08.500 & 50:56:41.01 & $24.85\pm0.02$ & $24.32\pm0.03$ \\ 
\hline
S41 & 4 &  85.5 & 262.1 & 09:22:12.214 & 50:56:37.22 & $23.26\pm0.01$ & $22.95\pm0.02$ \\ 
S42 & 4 & 588.9 & 323.0 & 09:22:12.322 & 50:57:27.53 & $24.52\pm0.01$ & $24.49\pm0.03$ \\ 
S43 & 4 & 296.7 & 300.6 & 09:22:12.392 & 50:56:58.40 & $23.90\pm0.01$ & $23.80\pm0.01$ \\ 
S44 & 4 & 132.5 & 309.9 & 09:22:12.663 & 50:56:42.30 & $25.08\pm0.02$ & $24.58\pm0.04$ \\ 
S45 & 4 & 202.5 & 371.4 & 09:22:13.233 & 50:56:49.80 & $24.41\pm0.01$ & $23.06\pm0.01$ \\ 
S46 & 4 & 502.3 & 421.4 & 09:22:13.443 & 50:57:19.95 & $23.95\pm0.01$ & $23.58\pm0.02$ \\ 
S47 & 4 & 154.1 & 389.7 & 09:22:13.476 & 50:56:45.21 & $24.79\pm0.01$ & $24.72\pm0.03$ \\ 
S48 & 4 & 257.2 & 420.9 & 09:22:13.695 & 50:56:55.69 & $22.80\pm0.01$ & $21.97\pm0.01$ \\ 
S49 & 4 & 229.0 & 439.2 & 09:22:13.916 & 50:56:53.09 & $24.83\pm0.02$ & $24.37\pm0.03$ \\ 
S50 & 4 & 684.1 & 510.6 & 09:22:14.185 & 50:57:38.74 & $24.76\pm0.01$ & $24.65\pm0.03$ \\ 
S51 & 4 & 738.3 & 543.7 & 09:22:14.473 & 50:57:44.37 & $24.76\pm0.01$ & $24.31\pm0.03$ \\ 
S52 & 4 & 595.7 & 535.8 & 09:22:14.543 & 50:57:30.30 & $26.11\pm0.05$ & $23.25\pm0.01$ \\ 
S53 & 4 & 507.6 & 545.1 & 09:22:14.733 & 50:57:21.69 & $24.94\pm0.02$ & $24.06\pm0.05$ \\ 
S54 & 4 & 595.3 & 562.4 & 09:22:14.821 & 50:57:30.52 & $24.66\pm0.01$ & $24.06\pm0.02$ \\ 
S55 & 4 & 722.2 & 577.1 & 09:22:14.838 & 50:57:43.11 & $24.57\pm0.01$ & $23.54\pm0.02$ \\ 
S56 & 4 & 752.0 & 587.9 & 09:22:14.919 & 50:57:46.12 & $24.44\pm0.01$ & $24.10\pm0.02$ \\ 
S57 & 4 & 157.4 & 539.7 & 09:22:15.042 & 50:56:47.04 & $23.71\pm0.01$ & $23.39\pm0.02$ \\ 
S58 & 4 & 131.8 & 652.5 & 09:22:16.243 & 50:56:45.68 & $25.90\pm0.03$ & $23.45\pm0.01$ \\ 
S59 & 4 & 662.6 & 737.1 & 09:22:16.565 & 50:57:38.77 & $23.54\pm0.01$ & $22.82\pm0.01$ \\ 
S60 & 4 & 481.2 & 718.5 & 09:22:16.569 & 50:57:20.77 & $24.85\pm0.01$ & $23.93\pm0.02$ \\ 
S61 & 4 & 459.2 & 747.9 & 09:22:16.897 & 50:57:18.89 & $23.49\pm0.01$ & $22.94\pm0.01$ \\ 
\enddata
\end{deluxetable}

\begin{deluxetable}{lcrrrrcccc}
\tablecaption{Variables discovered in \n2841}
\tablenum{4}
\tablewidth{0pt}
\tablehead{\multicolumn{2}{c}{ID\ \ Chip} & \colhead{x} & \colhead{y} & \multicolumn{2}{c}{R.A. (J2000.0) Dec.}
& \colhead{P (d)} & \colhead{V (mag)} & \colhead{I (mag)} & \colhead{}}
\startdata
C01 & 4 &  51.7 & 387.2 & 09:22:13.556 & +50:56:35.13& $13.9\pm0.6$  & $26.61\pm0.05$ & $25.64\pm0.11$ &   \\ 
C02 & 3 & 263.2 & 194.9 & 09:22:07.430 & +50:56:43.10& $16.1\pm0.5$  & $25.96\pm0.03$ & $25.05\pm0.06$ &   \\ 
C03 & 4 & 660.6 & 112.5 & 09:22:10.055 & +50:57:32.45& $16.5\pm0.2$  & $26.30\pm0.04$ & $25.06\pm0.07$ &   \\ 
C04 & 4 & 284.2 & 325.8 & 09:22:12.669 & +50:56:57.42& $18.5\pm0.5$  & $26.35\pm0.04$ & $25.19\pm0.05$ &   \\ 
C05 & 3 &  52.3 & 355.3 & 09:22:09.441 & +50:57:01.23& $18.9\pm0.6$  & $24.84\pm0.02$ & $23.95\pm0.05$ &   \\ 
C06 & 2 & 111.0 & 118.1 & 09:22:09.112 & +50:56:23.87& $20.0\pm0.3$  & $25.67\pm0.04$ & $24.92\pm0.04$ & * \\ 
C07 & 3 & 438.1 & 165.2 & 09:22:05.636 & +50:56:38.26& $21.0\pm0.4$  & $25.96\pm0.03$ & $24.59\pm0.11$ &   \\ 
C08 & 2 & 129.0 & 102.3 & 09:22:09.298 & +50:56:22.28& $21.1\pm1.1$  & $25.39\pm0.03$ & $24.38\pm0.04$ &   \\ 
C09 & 4 & 378.8 & 311.1 & 09:22:12.416 & +50:57:06.63& $21.5\pm0.5$  & $25.96\pm0.03$ & $24.73\pm0.09$ &   \\ 
C10 & 3 & 762.2 & 719.8 & 09:22:01.632 & +50:57:29.27& $21.7\pm1.0$  & $26.06\pm0.06$ & $25.15\pm0.05$ & * \\ 
C11 & 3 & 675.6 & 715.4 & 09:22:02.530 & +50:57:29.84& $21.8\pm0.7$  & $26.09\pm0.05$ & $24.96\pm0.14$ & * \\ 
C12 & 4 & 437.0 & 292.6 & 09:22:12.161 & +50:57:12.21& $22.5\pm0.3$  & $25.71\pm0.02$ & $24.77\pm0.03$ & * \\ 
C13 & 1 & 382.6 & 145.9 & 09:22:11.681 & +50:56:28.38& $23.0\pm0.8$  & $26.79\pm0.05$ & $25.43\pm0.08$ & * \\ 
C14 & 4 &  59.4 & 290.3 & 09:22:12.536 & +50:56:34.94& $23.2\pm0.4$  & $26.42\pm0.03$ & $25.32\pm0.05$ & * \\ 
C15 & 4 & 251.0 & 293.2 & 09:22:12.363 & +50:56:53.81& $23.2\pm0.3$  & $25.94\pm0.03$ & $24.89\pm0.08$ & * \\ 
C16 & 3 & 666.2 & 528.2 & 09:22:02.836 & +50:57:11.58& $26.5\pm0.8$  & $25.88\pm0.05$ & $24.89\pm0.06$ & * \\ 
C17 & 4 & 626.0 & 338.8 & 09:22:12.449 & +50:57:31.34& $27.3\pm1.0$  & $25.96\pm0.04$ & $24.97\pm0.05$ & * \\ 
C18 & 3 & 328.2 &  66.4 & 09:22:06.896 & +50:56:29.77& $27.8\pm0.6$  & $25.73\pm0.05$ & $24.70\pm0.05$ & * \\ 
C19 & 4 & 334.8 &  66.9 & 09:22:09.914 & +50:56:59.89& $29.0\pm0.9$  & $25.99\pm0.03$ & $24.93\pm0.05$ & * \\ 
C20 & 2 & 253.8 &  75.2 & 09:22:09.735 & +50:56:10.33& $30.2\pm0.5$  & $25.88\pm0.05$ & $24.90\pm0.05$ & * \\ 
C21 & 4 & 551.9 & 191.4 & 09:22:10.985 & +50:57:22.56& $32.2\pm0.7$  & $25.04\pm0.01$ & $24.01\pm0.03$ & * \\ 
C22 & 1 & 689.7 & 435.7 & 09:22:13.295 & +50:56:16.80& $33.2\pm0.8$  & $25.38\pm0.03$ & $24.24\pm0.02$ & * \\ 
C23 & 3 & 487.1 & 227.6 & 09:22:05.051 & +50:56:43.88& $34.6\pm1.4$  & $25.17\pm0.03$ & $24.31\pm0.04$ & * \\ 
C24 & 4 &  88.9 & 281.4 & 09:22:12.412 & +50:56:37.74& $36.6\pm1.8$  & $25.84\pm0.04$ & $24.66\pm0.03$ & * \\ 
C25 & 2 & 291.2 &  93.9 & 09:22:09.587 & +50:56:06.44& $39.5\pm0.8$  & $25.31\pm0.02$ & $24.26\pm0.05$ & * \\ 
C26 & 4 & 532.1 & 311.9 & 09:22:12.264 & +50:57:21.81& $40.9\pm0.5$  & $25.15\pm0.02$ & $24.06\pm0.05$ & * \\ 
\hline 
V01 & 2 & 218.2 & 384.2 & 09:22:06.469 & +50:56:10.35&    $\dots   $ & $24.75\pm0.04$ & $23.62\pm0.04$ &   \\ 
V02 & 3 & 349.3 & 157.6 & 09:22:06.573 & +50:56:38.48&    $\dots   $ & $24.80\pm0.05$ & $23.49\pm0.04$ &   \\ 
V03 & 1 & 460.2 & 188.6 & 09:22:12.073 & +50:56:26.84&    $\dots   $ & $25.24\pm0.04$ & $24.02\pm0.05$ &   \\ 
\enddata
\tablecomments{(*): Variable was used to construct the P-L relations shown in Figures 8 and 9.}
\end{deluxetable}

\clearpage

\begin{deluxetable}{lccccc}
\tablecaption{Individual V band photometric measurements}
\tablenum{5}
\tablewidth{0pt}
\tablehead{\colhead{MJD} & \multicolumn{5}{c}{Variables}}
\startdata
       &      C01       &      C02       &      C03       &       C04      &       C05      \\
\hline
603.36 & $26.28\pm0.16$ & $25.48\pm0.21$ & $25.74\pm0.12$ & $26.33\pm0.13$ & $24.55\pm0.04$ \\
603.38 & $26.50\pm0.28$ & $25.53\pm0.14$ & $25.66\pm0.16$ & $26.09\pm0.16$ & $24.60\pm0.06$ \\
605.30 & $26.91\pm0.34$ & $25.69\pm0.14$ & $26.06\pm0.20$ & $26.59\pm0.38$ & $24.66\pm0.13$ \\
605.32 & $26.98\pm0.23$ & $25.64\pm0.16$ & $26.07\pm0.17$ & $26.63\pm0.21$ & $24.60\pm0.09$ \\
607.25 &    $\dots$     & $26.13\pm0.20$ & $26.25\pm0.18$ & $26.60\pm0.24$ & $24.67\pm0.05$ \\
607.26 & $27.21\pm0.28$ & $26.02\pm0.25$ & $26.40\pm0.24$ & $26.74\pm0.24$ & $24.79\pm0.08$ \\
609.67 & $27.21\pm0.36$ & $26.20\pm0.20$ & $26.78\pm0.22$ & $26.43\pm0.22$ & $24.95\pm0.06$ \\
609.68 & $27.88\pm0.68$ & $25.85\pm0.18$ & $26.90\pm0.52$ & $27.11\pm0.30$ & $24.93\pm0.13$ \\
612.55 & $26.61\pm0.24$ &    $\dots$     & $26.80\pm0.27$ &    $\dots$     & $24.92\pm0.06$ \\
612.56 & $26.95\pm0.27$ & $26.16\pm0.22$ & $27.18\pm0.55$ & $26.84\pm0.19$ & $24.85\pm0.11$ \\
615.50 & $26.00\pm0.16$ & $26.34\pm0.26$ & $26.73\pm0.22$ & $26.27\pm0.23$ & $25.18\pm0.09$ \\
615.51 & $26.01\pm0.19$ &    $\dots$     & $27.16\pm0.46$ & $26.16\pm0.24$ & $25.06\pm0.15$ \\
619.39 & $26.64\pm0.22$ & $25.57\pm0.14$ & $25.57\pm0.11$ & $26.27\pm0.25$ & $24.94\pm0.06$ \\
619.40 & $26.78\pm0.24$ & $25.63\pm0.08$ & $25.77\pm0.19$ & $25.98\pm0.12$ & $25.10\pm0.13$ \\
623.61 & $27.49\pm0.38$ & $25.97\pm0.15$ & $26.36\pm0.16$ & $26.25\pm0.15$ & $24.55\pm0.05$ \\
623.62 & $26.76\pm0.24$ & $26.14\pm0.15$ & $26.48\pm0.31$ & $26.45\pm0.25$ & $24.77\pm0.06$ \\
629.31 & $26.18\pm0.22$ &    $\dots$     & $26.49\pm0.31$ & $26.54\pm0.26$ & $24.74\pm0.08$ \\
629.32 & $26.14\pm0.19$ & $26.14\pm0.23$ & $26.93\pm0.30$ & $26.93\pm0.31$ & $24.96\pm0.09$ \\
636.28 & $27.21\pm0.37$ & $25.69\pm0.13$ & $25.79\pm0.17$ & $26.16\pm0.18$ & $24.87\pm0.09$ \\
636.29 & $27.78\pm0.84$ &    $\dots$     & $25.77\pm0.15$ & $25.95\pm0.15$ & $25.00\pm0.08$ \\
643.25 & $26.27\pm0.13$ & $26.16\pm0.17$ & $27.08\pm0.35$ & $26.32\pm0.26$ & $24.68\pm0.10$ \\
643.27 & $26.03\pm0.12$ & $25.92\pm0.22$ & $27.14\pm0.36$ & $27.04\pm0.41$ & $24.65\pm0.11$ \\
653.57 & $27.16\pm0.52$ & $25.75\pm0.17$ & $25.86\pm0.11$ & $26.09\pm0.19$ & $25.01\pm0.12$ \\
653.59 & $27.23\pm0.66$ & $25.62\pm0.23$ & $25.76\pm0.20$ & $25.77\pm0.10$ & $25.09\pm0.17$ \\
\hline
       &      C06       &      C07       &      C08       &       C09      &       C10      \\
\hline
603.36 & $26.27\pm0.22$ & $25.81\pm0.15$ & $25.32\pm0.10$ &    $\dots$     & $27.21\pm0.52$ \\ 
603.38 & $26.19\pm0.17$ & $25.48\pm0.12$ & $25.19\pm0.14$ & $26.14\pm0.17$ & $27.24\pm0.39$ \\ 
605.30 & $26.44\pm0.35$ & $25.64\pm0.10$ & $25.39\pm0.14$ & $26.42\pm0.23$ & $26.30\pm0.23$ \\ 
605.32 & $26.29\pm0.16$ & $25.53\pm0.11$ & $25.28\pm0.07$ & $26.16\pm0.15$ & $26.34\pm0.43$ \\ 
607.25 & $25.18\pm0.10$ & $25.82\pm0.09$ & $25.33\pm0.12$ & $26.57\pm0.22$ & $25.65\pm0.16$ \\ 
607.26 & $25.20\pm0.11$ & $25.81\pm0.15$ & $25.37\pm0.07$ & $26.64\pm0.20$ & $25.50\pm0.20$ \\ 
609.67 & $25.25\pm0.06$ & $26.02\pm0.24$ & $25.31\pm0.16$ & $26.34\pm0.13$ & $25.86\pm0.15$ \\ 
609.68 & $25.24\pm0.10$ & $25.98\pm0.15$ &    $\dots$     & $26.18\pm0.17$ & $26.09\pm0.32$ \\ 
612.55 & $25.36\pm0.11$ & $26.12\pm0.25$ & $25.43\pm0.12$ & $25.86\pm0.13$ & $25.74\pm0.21$ \\ 
612.56 & $25.39\pm0.11$ & $26.33\pm0.24$ & $25.46\pm0.07$ & $26.12\pm0.12$ & $26.08\pm0.35$ \\ 
615.50 & $25.82\pm0.14$ & $26.39\pm0.26$ & $25.77\pm0.16$ & $25.38\pm0.09$ & $26.07\pm0.17$ \\ 
615.51 & $25.89\pm0.15$ & $26.32\pm0.23$ & $25.65\pm0.27$ & $25.59\pm0.11$ & $26.12\pm0.28$ \\ 
619.39 & $25.76\pm0.13$ & $26.49\pm0.25$ & $25.42\pm0.11$ & $25.83\pm0.12$ & $26.73\pm0.46$ \\ 
619.40 & $26.00\pm0.17$ & $26.64\pm0.24$ & $25.83\pm0.13$ & $25.98\pm0.10$ & $26.88\pm0.47$ \\ 
623.61 & $26.18\pm0.21$ & $25.71\pm0.10$ & $24.95\pm0.19$ & $26.01\pm0.12$ & $26.53\pm0.24$ \\ 
623.62 & $25.91\pm0.15$ & $25.84\pm0.14$ & $24.95\pm0.12$ & $26.00\pm0.14$ & $26.94\pm0.33$ \\ 
629.31 & $25.27\pm0.09$ & $25.55\pm0.12$ & $25.53\pm0.16$ & $26.46\pm0.25$ & $25.55\pm0.18$ \\ 
629.32 & $25.40\pm0.13$ & $25.72\pm0.16$ & $25.62\pm0.15$ & $26.44\pm0.18$ & $25.44\pm0.18$ \\ 
636.28 & $25.58\pm0.13$ & $26.27\pm0.20$ & $25.69\pm0.20$ & $25.40\pm0.10$ & $25.79\pm0.14$ \\ 
636.29 & $25.80\pm0.10$ & $26.00\pm0.13$ & $25.69\pm0.23$ & $25.38\pm0.08$ & $25.77\pm0.13$ \\ 
643.25 & $26.13\pm0.20$ & $26.09\pm0.20$ & $25.08\pm0.07$ & $25.86\pm0.16$ & $26.32\pm0.23$ \\ 
643.27 & $26.07\pm0.12$ & $26.40\pm0.58$ & $25.06\pm0.12$ & $25.84\pm0.14$ & $26.30\pm0.23$ \\ 
653.57 & $25.52\pm0.18$ & $25.96\pm0.13$ & $25.45\pm0.15$ & $26.65\pm0.36$ & $26.00\pm0.17$ \\ 
653.59 &    $\dots$     & $26.23\pm0.20$ &    $\dots$     & $26.43\pm0.23$ &   $\dots$      \\
\hline 
       &      C11       &      C12       &      C13       &       C14      &       C15      \\
\hline
603.36 &    $\dots$     & $25.87\pm0.15$ & $26.15\pm0.18$ & $25.95\pm0.13$ & $26.19\pm0.19$ \\ 
603.38 & $26.72\pm0.39$ & $26.07\pm0.20$ & $26.10\pm0.12$ & $25.92\pm0.10$ & $26.44\pm0.21$ \\ 
605.30 & $26.36\pm0.26$ & $25.90\pm0.12$ & $26.59\pm0.26$ & $26.17\pm0.21$ & $25.40\pm0.15$ \\ 
605.32 &    $\dots$     & $25.65\pm0.14$ & $26.40\pm0.17$ & $26.05\pm0.18$ & $25.42\pm0.16$ \\ 
607.25 & $25.33\pm0.11$ & $25.34\pm0.08$ & $26.43\pm0.22$ & $26.50\pm0.20$ & $25.30\pm0.08$ \\ 
607.26 & $25.48\pm0.20$ & $25.38\pm0.07$ & $26.61\pm0.19$ & $26.44\pm0.26$ & $25.30\pm0.08$ \\ 
609.67 & $25.68\pm0.16$ & $25.39\pm0.08$ & $26.71\pm0.26$ & $26.54\pm0.22$ & $25.56\pm0.10$ \\ 
609.68 & $25.85\pm0.28$ & $25.32\pm0.08$ & $26.97\pm0.37$ & $26.64\pm0.25$ & $25.64\pm0.15$ \\ 
612.55 & $25.93\pm0.14$ & $25.70\pm0.11$ & $27.12\pm0.23$ & $26.90\pm0.29$ & $25.95\pm0.13$ \\ 
612.56 & $26.25\pm0.27$ & $25.55\pm0.13$ & $27.20\pm0.34$ & $26.66\pm0.20$ & $25.94\pm0.15$ \\ 
615.50 & $26.33\pm0.23$ & $25.64\pm0.20$ & $27.08\pm0.45$ &    $\dots$     & $25.87\pm0.16$ \\ 
615.51 &    $\dots$     & $25.65\pm0.14$ & $26.83\pm0.26$ & $27.34\pm0.35$ & $26.20\pm0.16$ \\ 
619.39 & $26.60\pm0.31$ & $25.82\pm0.13$ &    $\dots$     & $26.91\pm0.20$ &   $\dots$      \\ 
619.40 &    $\dots$     & $25.72\pm0.41$ &    $\dots$     &    $\dots$     & $26.15\pm0.17$ \\ 
623.61 & $26.95\pm0.34$ & $26.04\pm0.12$ & $26.70\pm0.24$ & $25.97\pm0.19$ & $26.80\pm0.34$ \\ 
623.62 & $26.77\pm0.38$ & $26.11\pm0.14$ & $27.26\pm0.48$ & $25.65\pm0.15$ & $26.33\pm0.12$ \\ 
629.31 & $25.23\pm0.13$ & $25.47\pm0.07$ & $26.30\pm0.23$ & $26.22\pm0.12$ & $25.42\pm0.14$ \\ 
629.32 & $25.57\pm0.14$ & $25.60\pm0.19$ & $26.61\pm0.27$ & $26.24\pm0.14$ & $25.41\pm0.06$ \\ 
636.28 & $25.89\pm0.19$ & $25.77\pm0.09$ & $27.49\pm0.40$ & $27.19\pm0.50$ & $25.55\pm0.22$ \\ 
636.29 & $26.48\pm0.35$ & $25.71\pm0.12$ & $27.05\pm0.44$ & $27.17\pm0.29$ & $25.94\pm0.12$ \\ 
643.25 & $26.75\pm0.27$ & $26.08\pm0.17$ & $27.40\pm0.46$ & $27.16\pm0.51$ & $26.30\pm0.22$ \\ 
643.27 & $26.22\pm0.27$ & $25.80\pm0.12$ & $27.22\pm0.34$ & $26.84\pm0.22$ & $26.53\pm0.22$ \\ 
653.57 & $25.66\pm0.28$ & $25.23\pm0.10$ & $26.34\pm0.24$ & $26.15\pm0.19$ & $25.31\pm0.10$ \\ 
653.59 & $25.77\pm0.18$ & $25.36\pm0.09$ & $26.82\pm0.23$ & $26.25\pm0.17$ & $25.51\pm0.11$ \\ 
\hline
       &      C16       &      C17       &      C18       &       C19      &       C20      \\
\hline
603.36 & $26.22\pm0.16$ & $26.14\pm0.23$ & $25.68\pm0.38$ & $25.58\pm0.12$ & $25.84\pm0.15$ \\ 
603.38 & $26.05\pm0.23$ & $26.37\pm0.29$ & $25.63\pm0.14$ & $25.76\pm0.19$ & $25.78\pm0.11$ \\ 
605.30 & $26.67\pm0.35$ & $26.25\pm0.28$ & $26.19\pm0.17$ & $25.66\pm0.14$ & $26.00\pm0.10$ \\ 
605.32 & $26.28\pm0.32$ & $26.22\pm0.15$ & $26.05\pm0.31$ & $25.74\pm0.18$ & $26.05\pm0.20$ \\ 
607.25 & $26.30\pm0.19$ & $26.39\pm0.30$ & $26.36\pm0.23$ & $25.60\pm0.06$ & $26.40\pm0.18$ \\ 
607.26 & $26.91\pm0.39$ & $26.34\pm0.29$ & $25.95\pm0.25$ & $25.98\pm0.16$ & $25.99\pm0.11$ \\ 
609.67 & $26.43\pm0.27$ & $26.46\pm0.18$ & $26.36\pm0.16$ &    $\dots$     & $26.06\pm0.25$ \\ 
609.68 & $25.96\pm0.15$ & $26.58\pm0.23$ & $26.46\pm0.23$ & $26.05\pm0.35$ & $26.05\pm0.18$ \\ 
612.55 & $25.47\pm0.10$ & $26.04\pm0.11$ & $26.27\pm0.13$ & $26.12\pm0.34$ & $26.59\pm0.22$ \\ 
612.56 & $25.50\pm0.18$ & $26.02\pm0.16$ & $26.67\pm0.34$ & $26.09\pm0.14$ & $26.17\pm0.10$ \\ 
615.50 & $25.52\pm0.13$ &    $\dots$     & $25.77\pm0.24$ &    $\dots$     & $26.72\pm0.23$ \\ 
615.51 & $25.48\pm0.15$ & $26.31\pm0.27$ & $25.78\pm0.15$ & $26.46\pm0.39$ & $26.09\pm0.14$ \\ 
619.39 & $25.70\pm0.20$ & $25.28\pm0.18$ & $25.00\pm0.07$ & $26.59\pm0.30$ & $26.53\pm0.23$ \\ 
619.40 & $25.62\pm0.09$ & $25.52\pm0.11$ & $25.16\pm0.12$ & $26.48\pm0.21$ & $26.84\pm0.28$ \\ 
623.61 & $25.79\pm0.17$ & $25.88\pm0.10$ & $25.55\pm0.14$ & $26.62\pm0.21$ & $25.46\pm0.07$ \\ 
623.62 & $26.00\pm0.15$ &    $\dots$     & $25.52\pm0.09$ & $26.58\pm0.29$ & $25.50\pm0.12$ \\ 
629.31 & $26.68\pm0.32$ & $26.24\pm0.22$ & $26.09\pm0.18$ & $25.50\pm0.12$ & $25.45\pm0.11$ \\ 
629.32 & $26.91\pm0.47$ & $26.19\pm0.32$ & $25.80\pm0.21$ & $25.57\pm0.11$ &   $\dots$      \\ 
636.28 & $26.20\pm0.15$ & $26.43\pm0.27$ & $26.80\pm0.41$ &    $\dots$     & $25.86\pm0.11$ \\ 
636.29 & $26.13\pm0.18$ & $26.49\pm0.27$ & $26.11\pm0.13$ & $25.88\pm0.10$ & $26.29\pm0.19$ \\ 
643.25 &    $\dots$     & $25.77\pm0.15$ & $26.65\pm0.45$ & $26.49\pm0.17$ & $26.45\pm0.22$ \\ 
643.27 &    $\dots$     & $26.00\pm0.25$ & $26.10\pm0.18$ & $26.08\pm0.16$ & $26.18\pm0.21$ \\ 
653.57 & $26.22\pm0.17$ & $26.07\pm0.18$ & $25.61\pm0.13$ & $26.30\pm0.25$ & $25.45\pm0.14$ \\ 
653.59 & $26.57\pm0.26$ & $26.43\pm0.42$ &    $\dots$     & $26.65\pm0.34$ &   $\dots$      \\
\hline
       &      C21       &      C22       &      C23       &       C24      &       C25      \\
\hline
603.36 & $25.29\pm0.09$ & $25.02\pm0.07$ & $24.62\pm0.09$ & $25.82\pm0.16$ & $25.65\pm0.12$ \\ 
603.38 & $25.29\pm0.15$ & $24.71\pm0.09$ & $24.48\pm0.22$ & $25.61\pm0.17$ & $25.36\pm0.14$ \\ 
605.30 & $25.19\pm0.10$ & $24.74\pm0.34$ & $24.76\pm0.07$ & $25.94\pm0.13$ & $25.69\pm0.09$ \\ 
605.32 & $25.02\pm0.12$ & $25.13\pm0.09$ &    $\dots$     & $25.89\pm0.14$ & $25.87\pm0.22$ \\ 
607.25 & $24.97\pm0.08$ & $25.10\pm0.09$ & $25.15\pm0.08$ & $26.01\pm0.17$ & $25.83\pm0.18$ \\ 
607.26 & $24.70\pm0.11$ & $25.14\pm0.09$ & $24.73\pm0.08$ & $25.82\pm0.19$ & $25.74\pm0.13$ \\ 
609.67 & $24.72\pm0.06$ & $25.32\pm0.08$ & $25.13\pm0.08$ & $26.28\pm0.13$ &   $\dots$      \\ 
609.68 & $24.69\pm0.08$ & $25.32\pm0.10$ &    $\dots$     & $25.82\pm0.19$ & $25.66\pm0.13$ \\ 
612.55 & $24.81\pm0.07$ & $25.26\pm0.09$ & $25.05\pm0.09$ & $26.00\pm0.13$ & $25.72\pm0.12$ \\ 
612.56 & $24.75\pm0.06$ & $25.18\pm0.08$ & $25.03\pm0.11$ & $25.83\pm0.11$ & $25.58\pm0.13$ \\ 
615.50 & $24.81\pm0.08$ & $25.79\pm0.14$ & $25.22\pm0.09$ & $26.01\pm0.16$ & $25.32\pm0.13$ \\ 
615.51 & $24.88\pm0.11$ & $25.85\pm0.22$ & $25.39\pm0.12$ & $26.48\pm0.24$ & $25.32\pm0.10$ \\ 
619.39 &    $\dots$     &    $\dots$     & $25.74\pm0.17$ & $26.31\pm0.12$ & $24.89\pm0.07$ \\ 
619.40 & $25.00\pm0.07$ & $26.01\pm0.19$ & $25.58\pm0.13$ & $26.00\pm0.19$ & $24.78\pm0.11$ \\ 
623.61 & $25.10\pm0.07$ & $25.97\pm0.16$ & $25.54\pm0.09$ & $26.40\pm0.15$ & $25.03\pm0.09$ \\ 
623.62 & $25.17\pm0.10$ & $25.60\pm0.13$ & $25.30\pm0.06$ & $26.50\pm0.25$ & $24.95\pm0.08$ \\ 
629.31 & $25.39\pm0.15$ &    $\dots$     & $25.78\pm0.22$ & $25.39\pm0.09$ & $25.19\pm0.10$ \\ 
629.32 & $25.34\pm0.12$ & $25.66\pm0.14$ & $25.75\pm0.18$ & $25.29\pm0.11$ & $25.13\pm0.09$ \\ 
636.28 & $25.31\pm0.15$ & $24.81\pm0.10$ & $24.66\pm0.08$ & $25.51\pm0.11$ & $25.55\pm0.08$ \\ 
636.29 & $25.25\pm0.10$ & $24.83\pm0.08$ &    $\dots$     & $25.70\pm0.17$ & $25.42\pm0.09$ \\ 
643.25 & $24.75\pm0.09$ & $25.30\pm0.10$ & $25.05\pm0.13$ & $25.96\pm0.14$ & $25.62\pm0.11$ \\ 
643.27 & $24.68\pm0.07$ & $25.29\pm0.09$ & $24.91\pm0.07$ & $25.98\pm0.14$ & $25.46\pm0.17$ \\ 
653.57 & $25.07\pm0.10$ & $25.76\pm0.14$ &    $\dots$     & $26.60\pm0.34$ & $25.61\pm0.12$ \\ 
653.59 &    $\dots$     & $25.73\pm0.13$ & $25.45\pm0.12$ & $26.13\pm0.21$ &   $\dots$      \\
\hline
       &      C26       &      V01       &      V02       &       V03      &                \\
\hline
603.36 & $25.73\pm0.09$ & $24.87\pm0.08$ & $24.84\pm0.05$ & $25.49\pm0.09$ &                \\ 
603.38 &    $\dots$     & $24.86\pm0.06$ & $24.77\pm0.11$ & $25.41\pm0.13$ &                \\ 
605.30 & $25.51\pm0.10$ & $24.87\pm0.05$ & $24.70\pm0.04$ & $25.34\pm0.11$ &                \\ 
605.32 & $25.67\pm0.10$ & $24.87\pm0.05$ & $24.68\pm0.07$ & $25.38\pm0.12$ &                \\ 
607.25 &    $\dots$     &    $\dots$     & $24.50\pm0.06$ & $25.45\pm0.13$ &                \\ 
607.26 & $25.60\pm0.14$ & $24.86\pm0.07$ & $24.56\pm0.07$ & $25.40\pm0.13$ &                \\ 
609.67 & $25.48\pm0.09$ & $24.93\pm0.10$ & $24.58\pm0.07$ & $25.35\pm0.11$ &                \\ 
609.68 & $25.42\pm0.09$ & $24.89\pm0.08$ & $24.59\pm0.07$ & $25.39\pm0.08$ &                \\ 
612.55 & $24.80\pm0.05$ & $24.92\pm0.05$ & $24.51\pm0.05$ & $25.44\pm0.09$ &                \\ 
612.56 & $24.94\pm0.05$ &    $\dots$     & $24.58\pm0.06$ & $25.28\pm0.10$ &                \\ 
615.50 & $24.61\pm0.08$ & $24.98\pm0.08$ & $24.62\pm0.06$ & $25.40\pm0.16$ &                \\ 
615.51 & $24.71\pm0.04$ & $25.01\pm0.10$ & $24.55\pm0.09$ & $25.36\pm0.10$ &                \\ 
619.39 & $24.75\pm0.04$ & $24.85\pm0.06$ & $24.71\pm0.10$ &    $\dots$     &                \\ 
619.40 & $24.88\pm0.05$ & $24.92\pm0.07$ & $24.80\pm0.10$ & $25.15\pm0.10$ &                \\ 
623.61 & $25.00\pm0.07$ & $24.78\pm0.05$ & $24.84\pm0.09$ & $24.98\pm0.10$ &                \\ 
623.62 & $25.03\pm0.06$ & $24.82\pm0.07$ & $24.86\pm0.12$ &    $\dots$     &                \\ 
629.31 & $25.23\pm0.12$ & $24.69\pm0.04$ & $24.84\pm0.06$ & $24.83\pm0.09$ &                \\ 
629.32 & $25.07\pm0.08$ & $24.74\pm0.10$ &    $\dots$     & $24.87\pm0.09$ &                \\ 
636.28 & $25.48\pm0.06$ & $24.55\pm0.06$ & $25.06\pm0.07$ & $25.08\pm0.07$ &                \\ 
636.29 & $25.39\pm0.07$ & $24.44\pm0.04$ & $25.02\pm0.08$ & $24.98\pm0.10$ &                \\ 
643.25 & $25.70\pm0.18$ & $24.43\pm0.05$ & $25.23\pm0.10$ & $24.79\pm0.17$ &                \\ 
643.27 & $25.55\pm0.05$ & $24.34\pm0.06$ & $24.93\pm0.08$ & $25.21\pm0.10$ &                \\ 
653.57 & $24.86\pm0.06$ & $24.41\pm0.09$ & $25.36\pm0.08$ & $25.38\pm0.11$ &                \\ 
653.59 &    $\dots$     &    $\dots$     & $25.20\pm0.06$ & $25.32\pm0.11$ &                \\ 
\enddata
\end{deluxetable}

\begin{deluxetable}{lccccc}
\tablecaption{Individual I band photometric measurements}
\tablenum{6}
\tablewidth{0pt}
\tablehead{\colhead{MJD} & \multicolumn{5}{c}{Variables}}
\startdata
       &      C01       &      C02       &      C03       &       C04      &       C05      \\
\hline
603.42 & $25.91\pm0.35$ & $24.90\pm0.19$ & $24.97\pm0.34$ & $25.35\pm0.23$ & $23.88\pm0.13$ \\ 
603.43 & $25.82\pm0.27$ & $24.73\pm0.12$ & $25.11\pm0.31$ & $25.39\pm0.20$ & $23.87\pm0.14$ \\ 
609.73 & $25.54\pm0.25$ & $24.95\pm0.31$ & $25.14\pm0.21$ & $25.50\pm0.26$ & $23.86\pm0.09$ \\ 
609.74 & $26.00\pm0.60$ & $25.04\pm0.21$ & $25.04\pm0.19$ & $25.51\pm0.22$ & $24.07\pm0.14$ \\ 
619.45 & $26.16\pm0.48$ & $24.80\pm0.46$ & $24.84\pm0.22$ & $25.12\pm0.17$ & $24.35\pm0.14$ \\ 
619.47 &    $\dots$     &    $\dots$     & $24.69\pm0.16$ & $25.00\pm0.14$ & $24.07\pm0.16$ \\ 
629.37 & $25.54\pm0.31$ & $25.61\pm0.43$ & $25.31\pm0.29$ & $25.27\pm0.16$ & $23.96\pm0.13$ \\ 
629.39 & $25.10\pm0.14$ & $25.52\pm0.28$ &    $\dots$     & $25.32\pm0.30$ & $24.16\pm0.15$ \\ 
643.32 & $25.50\pm0.24$ & $25.13\pm0.36$ & $25.02\pm0.16$ & $25.09\pm0.12$ & $23.61\pm0.09$ \\ 
643.33 & $25.38\pm0.20$ & $25.43\pm0.34$ & $25.44\pm0.41$ & $25.19\pm0.21$ & $23.75\pm0.10$ \\ 
\hline
       &      C06       &      C07       &      C08       &       C09      &       C10      \\
\hline
603.42 & $25.41\pm0.24$ & $24.59\pm0.15$ & $24.29\pm0.38$ & $25.34\pm0.26$ & $25.58\pm0.26$ \\ 
603.43 & $25.09\pm0.24$ &    $\dots$     & $24.39\pm0.16$ & $24.89\pm0.16$ & $25.49\pm0.37$ \\ 
609.73 & $24.59\pm0.12$ & $24.73\pm0.28$ & $24.22\pm0.15$ & $24.95\pm0.18$ & $25.13\pm0.31$ \\ 
609.74 & $24.69\pm0.17$ & $24.45\pm0.12$ & $24.35\pm0.12$ & $24.78\pm0.08$ & $24.97\pm0.19$ \\ 
619.45 & $25.08\pm0.21$ & $25.21\pm0.36$ & $24.50\pm0.22$ & $24.75\pm0.15$ & $25.23\pm0.25$ \\ 
619.47 & $25.22\pm0.34$ & $24.66\pm0.15$ & $24.64\pm0.19$ & $24.77\pm0.14$ & $25.13\pm0.28$ \\ 
629.37 & $24.57\pm0.09$ & $24.32\pm0.14$ & $24.28\pm0.11$ & $24.94\pm1.04$ & $24.85\pm0.18$ \\ 
629.39 & $24.64\pm0.15$ & $24.83\pm0.15$ & $24.56\pm0.24$ & $25.57\pm0.49$ & $24.99\pm0.22$ \\ 
643.32 & $25.42\pm0.23$ & $24.55\pm0.12$ & $24.23\pm0.14$ & $25.05\pm0.19$ & $25.23\pm0.21$ \\ 
643.33 & $25.32\pm0.23$ & $24.06\pm0.47$ & $24.23\pm0.16$ & $24.59\pm0.09$ &    $\dots$     \\ 
\hline
       &      C11       &      C12       &      C13       &       C14      &       C15      \\
\hline
603.42 &    $\dots$     & $24.99\pm0.15$ & $25.13\pm0.20$ & $24.87\pm0.18$ & $25.52\pm0.28$ \\ 
603.43 &    $\dots$     & $24.91\pm0.15$ & $25.23\pm0.22$ & $25.03\pm0.13$ & $25.50\pm0.40$ \\ 
609.73 & $25.12\pm0.37$ &    $\dots$     & $25.05\pm0.18$ & $25.30\pm0.20$ & $24.64\pm0.15$ \\ 
609.74 & $24.81\pm0.23$ & $24.80\pm0.19$ & $25.54\pm0.24$ & $25.12\pm0.18$ & $24.78\pm0.16$ \\ 
619.45 &    $\dots$     & $25.00\pm0.16$ &    $\dots$     &    $\dots$     & $24.90\pm0.18$ \\ 
619.47 &    $\dots$     & $24.83\pm0.12$ &    $\dots$     &    $\dots$     & $25.55\pm0.28$ \\ 
629.37 & $24.57\pm0.19$ & $24.60\pm0.11$ & $25.26\pm0.24$ &    $\dots$     & $24.43\pm0.10$ \\ 
629.39 & $24.72\pm0.12$ & $24.72\pm0.14$ &    $\dots$     & $25.27\pm0.18$ & $24.70\pm0.12$ \\ 
643.32 & $24.73\pm0.27$ & $25.02\pm0.30$ &    $\dots$     & $25.80\pm0.35$ & $25.54\pm0.25$ \\ 
643.33 & $25.21\pm0.25$ & $24.83\pm0.13$ &    $\dots$     &    $\dots$     & $25.18\pm0.21$ \\ 
\hline
       &      C16       &      C17       &      C18       &       C19      &       C20      \\
\hline
603.42 & $25.25\pm0.48$ & $25.21\pm0.24$ & $24.87\pm0.21$ & $24.78\pm0.10$ & $24.56\pm0.15$ \\ 
603.43 & $25.34\pm0.31$ & $25.13\pm0.27$ & $24.58\pm0.13$ & $24.66\pm0.15$ & $24.67\pm0.16$ \\ 
609.73 & $25.54\pm0.39$ & $25.44\pm0.21$ & $25.17\pm0.41$ & $24.68\pm0.24$ & $25.26\pm0.16$ \\ 
609.74 & $25.27\pm0.43$ & $25.33\pm0.33$ & $24.98\pm0.18$ & $24.58\pm0.17$ & $25.23\pm0.13$ \\ 
619.45 & $24.95\pm0.21$ & $24.67\pm0.12$ & $24.70\pm0.15$ & $25.13\pm0.33$ & $25.15\pm0.21$ \\ 
619.47 & $24.72\pm0.17$ & $24.99\pm0.28$ & $24.28\pm0.10$ & $25.52\pm0.29$ &    $\dots$     \\ 
629.37 &    $\dots$     & $25.02\pm0.21$ & $24.84\pm0.16$ & $24.52\pm0.14$ & $24.72\pm0.20$ \\ 
629.39 & $24.99\pm0.19$ & $24.76\pm0.17$ & $24.67\pm0.10$ & $24.80\pm0.14$ & $24.60\pm0.10$ \\ 
643.32 & $24.60\pm0.19$ & $25.05\pm0.34$ & $24.79\pm0.21$ & $24.93\pm0.14$ &    $\dots$     \\ 
643.33 & $24.53\pm0.17$ & $24.76\pm0.22$ & $24.78\pm0.16$ & $25.11\pm0.26$ & $25.09\pm0.17$ \\ 
\hline
       &      C21       &      C22       &      C23       &       C24      &       C25      \\
\hline
603.42 &    $\dots$     & $24.05\pm0.10$ & $24.08\pm0.10$ & $24.47\pm0.14$ &    $\dots$     \\ 
603.43 & $24.18\pm0.10$ & $23.94\pm0.12$ & $24.10\pm0.10$ & $24.55\pm0.20$ & $24.54\pm0.17$ \\ 
609.73 & $23.72\pm0.07$ & $24.05\pm0.08$ & $24.20\pm0.21$ & $24.57\pm0.12$ & $24.66\pm0.16$ \\ 
609.74 & $23.75\pm0.06$ & $24.11\pm0.12$ & $24.26\pm0.10$ & $24.73\pm0.21$ & $24.59\pm0.12$ \\ 
619.45 & $24.00\pm0.11$ & $24.34\pm0.11$ & $24.59\pm0.17$ & $24.87\pm0.20$ & $24.02\pm0.07$ \\ 
619.47 & $24.07\pm0.09$ & $24.29\pm0.11$ & $24.59\pm0.19$ & $25.04\pm0.21$ & $23.99\pm0.10$ \\ 
629.37 & $24.19\pm0.11$ & $24.48\pm0.15$ & $24.64\pm0.19$ & $24.68\pm0.14$ & $24.12\pm0.10$ \\ 
629.39 & $24.34\pm0.18$ & $24.56\pm0.14$ & $24.72\pm0.15$ & $24.37\pm0.11$ & $23.98\pm0.08$ \\ 
643.32 & $23.91\pm0.07$ &    $\dots$     & $23.97\pm0.07$ & $24.78\pm0.14$ & $24.77\pm0.20$ \\ 
643.33 & $23.87\pm0.09$ & $24.01\pm0.08$ & $24.15\pm0.14$ & $24.60\pm0.14$ & $24.66\pm0.18$ \\ 
\hline
       &      C26       &      V01       &      V02       &       V03      &                \\
\hline
603.42 & $24.50\pm0.16$ & $23.60\pm0.06$ & $23.43\pm0.06$ & $24.19\pm0.14$ &                \\ 
603.43 & $24.78\pm0.21$ & $23.73\pm0.09$ & $23.49\pm0.04$ & $24.03\pm0.09$ &                \\ 
609.73 & $24.12\pm0.08$ & $23.75\pm0.27$ & $23.29\pm0.23$ & $24.14\pm0.13$ &                \\ 
609.74 & $24.35\pm0.18$ & $23.77\pm0.05$ & $23.42\pm0.08$ & $24.03\pm0.13$ &                \\ 
619.45 & $23.89\pm0.09$ & $23.72\pm0.06$ & $23.37\pm0.04$ & $24.16\pm0.09$ &                \\ 
619.47 & $23.88\pm0.10$ & $23.70\pm0.09$ & $23.48\pm0.06$ & $24.15\pm0.11$ &                \\ 
629.37 & $24.01\pm0.10$ & $23.56\pm0.06$ & $23.53\pm0.07$ & $23.77\pm0.09$ &                \\ 
629.39 & $23.92\pm0.09$ & $23.48\pm0.06$ & $23.58\pm0.08$ & $23.83\pm0.08$ &                \\ 
643.32 & $24.26\pm0.11$ & $23.49\pm0.04$ & $23.69\pm0.07$ &    $\dots$     &                \\ 
643.33 & $24.43\pm0.11$ & $23.44\pm0.09$ & $23.65\pm0.06$ & $23.84\pm0.09$ &                \\ 
\enddata
\end{deluxetable}

\clearpage

\begin{deluxetable}{lr}
\tablecaption{Error Budget for the Cepheid Distance to \n2841}
\tablenum{7}
\tablewidth{0pt}
\tablehead{\colhead{Item} & \colhead{Value (mag)}}
\startdata
Dispersion in $\mu_0$     & $\pm0.06$ \\
Metallicity correction    & $\pm0.15$ \\
LMC distance              & $\pm0.10$ \\
WFPC2 calibration         & $\pm0.07$ \\
Blending                  & $+0.10$   \\\hline
Total                     & $+0.23, -0.20$\\
\enddata
\end{deluxetable}

\begin{deluxetable}{lcllcccccc}
\tablecaption{Galaxies with Cepheid Distances in the Leo Cloud \& Spur}
\tablenum{8}
\tablewidth{0pt}
\tablehead{\colhead{Name} & \colhead{Cepheid} & \colhead{R.A.} & \colhead{Dec.}
& \colhead{$cz$} & \multicolumn{5}{c} {Super-galactic coordinates}\\
\colhead{} & \colhead{distance} & \multicolumn{2}{c}{(J2000.0)} & \colhead{km/s} & \colhead{Long.} & \colhead{Lat.} &
\colhead{X} & \colhead{Y} & \colhead{Z}\\
\colhead{} & \colhead{(Mpc)} & \colhead{} & \colhead{} & \colhead{} & \multicolumn{2}{c}{($^{\circ}$)} &
\multicolumn{3}{c}{(Mpc)}}
\startdata
NGC$\,$2541        & 11.2 & 08:14:40.2 & +49:03:42 & $559\pm1$ &  40.5 & -22.9 &  7.8 &  6.7 & -4.4 \\
{\bf NGC$\,$2841}  & 14.1 & 09:22:02.7 & +50:58:36 & $638\pm3$ &  49.6 & -16.0 &  8.7 & 10.3 & -3.9 \\
NGC$\,$3198        & 13.8 & 10:19:54.9 & +45:32:59 & $663\pm4$ &  60.6 & -13.3 &  6.6 & 11.7 & -3.2 \\
NGC$\,$3319        & 13.3 & 10:39:08.8 & +41:41:16 & $739\pm1$ &  66.0 & -12.8 &  5.3 & 11.8 & -2.9 \\
NGC$\,$3351        & 10.0 & 10:43:57.8 & +11:42:14 & $778\pm4$ &  94.1 & -27.1 & -0.6 &  8.9 & -4.6 \\
NGC$\,$3368        & 10.5 & 10:46:45.7 & +11:49:12 & $897\pm4$ &  94.3 & -26.4 & -0.7 &  9.4 & -4.7 \\
NGC$\,$3621        &  6.6 & 11:18:16.0 &\m32:48:42 & $727\pm5$ & 145.7 & -28.5 & -4.8 &  3.3 & -3.2 \\
NGC$\,$3627        & 10.1 & 11:20;15.0 & +12:59:30 & $727\pm3$ &  96.6 & -18.4 & -1.1 &  9.5 & -3.2 \\
\enddata
\end{deluxetable}

\begin{thebibliography}{}
\bibitem[Biretta et al.\,(2000)]{bir00} Biretta, J.A., et al.~2000,
{\it Wide Field and Planetary Camera 2 Instrument Handbook, Version
5.0} (Baltimore: STScI)

\bibitem[Bothun et al.\,(2001)]{bo01} Bothun, G.D., et al.~2001, \apj, in
preparation

\bibitem[Bresolin et al.\,(1999)]{bre99} Bresolin, F., et al.~1999, \apj, 510,
104

\bibitem[de Jong \& Lacey\,(2000)]{djl00} de Jong, R.S. \& Lacey, C.~2000,
\apj, 545, 781

\bibitem[de Vaucouleurs et al.\,(1991)]{RC3} de Vaucouleurs, G., et al.~1991,
{\it Third Reference Catalogue of Bright Galaxies} (Berlin: Springer-Verlag)

\bibitem[Freedman et al.\,(2001)]{fre01} Freedman, W.L., et al.~2001,
\apj, 553, 47

\bibitem[Garnavich et al.\,(2001)]{ga01} Garnavich, P.M., et al.~2001, \apj,
submitted

\bibitem[Hill et al.\,(1998)]{hil98} Hill, R.J., et al.~1998, \apj, 496, 648

\bibitem[Holtzman et al.\,(1995a)]{hol95a} Holtzman, J.A., et al.~1995a,
\pasp, 107, 156

\bibitem[Holtzman et al.\,(1995b)]{hol95b} Holtzman, J.A., et al.~1995b,
\pasp, 107, 1065

\bibitem[Kaluzny et al.\,(1998)]{ka98} Kaluzny, J., et al.~1998, \aj, 115,
1016

\bibitem[Kennicutt et al.\,(1998)]{ken98} Kennicutt, R.C., et al.~1998, \apj,
498, 181

\bibitem[Kochanek\,(1997)]{ko97} Kochanek, C.S.~1997, \apj, 491, 13

\bibitem[Madore\,(1982)]{ma82} Madore, B.F.~1982, \apj, 253, 575

\bibitem[Macri et al.\,(2000)]{ma00} Macri, L.M., et al.~2000, \apjs, 128, 461

\bibitem[Macri et al.\,(2001)]{ma01} Macri, L.M., et al.~2001, \apj, 548, xxx

\bibitem[Mochejska et al.\,(2001)]{mo01} Mochejska, B.J., et al.~2001, \aj, in
preparation

\bibitem[Rothberg et al.\,(2000)]{ro00} Rothberg, B., et al.~2000, \apj, 533,
781

\bibitem[Sakai et al.\,(2000)]{szk00} Sakai, S., et al.~2000, \aj, 119, 1197

\bibitem[Sandage \& Tammann\,(1981)]{RSA} Sandage, A. \& Tammann, G.~1981,
{\it Revised Shapley-Ames Catalog of Bright Galaxies} (Washington: Carnegie)

\bibitem[Sasselov et al.\,(1997)]{sa97} Sasselov, D.D., et al.~1997, \aap,
324, 471

\bibitem[Stetson\,(1990)]{ste90} Stetson, P.B.~1990, \pasp, 102, 932

\bibitem[Stetson\,(1994)]{ste94} Stetson, P.B.~1994, \pasp, 106, 250

\bibitem[Stetson\,(1996)]{ste96} Stetson, P.B.~1996, \pasp, 108, 851

\bibitem[Stetson\,(1998)]{ste98} Stetson, P.B.~1998, \pasp, 110, 1448

\bibitem[Tully \& Fisher\,(1987)]{NBA} Tully, R.B. \& Fisher, J.R.~1987, {\it
Nearby Galaxies Atlas} (Cambridge; New York: Cambridge University Press)

\bibitem[Tully\,(1988)]{NBG} Tully, R.B.~1988, {\it Nearby Galaxies Catalog}
(Cambridge; New York: Cambridge University Press)

\bibitem[Udalski et al.\,(1999)]{ud99} Udalski, A., et al.~1999, Ac.A., 49,
201

\bibitem[Zaritsky et al.\,(1994)]{zkh94} Zaritsky, D., et al.~1994, \apj, 420,
87

\end{thebibliography}
\end{document}